\documentclass[12pt,preprint]{aastex}
\def\plotfiddle#1#2#3#4#5#6#7{\centering \leavevmode
\vbox to#2{\rule{0pt}{#2}}
\includegraphics{#1}}

\slugcomment{Accepted for Publication in the Astrophysical Journal}

\usepackage{color}

\shorttitle{Extended Br$\gamma$ Emission in YSOs}
\shortauthors{Beck, Bary \& McGregor}

\begin{document}


\title{Spatially Extended Brackett Gamma Emission in the Environments of Young Stars\altaffilmark{1}}


\author{Tracy L. Beck}
\affil{Space Telescope Science Institute, 3700 San Martin Dr., Baltimore, MD 21218}
\email{tbeck@stsci.edu} 

\author{Jeffery S. Bary}
\affil{Colgate University, Department of Physics \& Astronomy,  13 Oak Drive, Hamilton, NY  13346}
\email{jbary@colgate.edu} 

\and 

\author{Peter J. McGregor}
\affil{Research School of Astronomy \& Astrophysics, Australian National University, Cotter Road, Weston, ACT 2611, Australia}
\email{peter@mso.anu.edu.au}

\altaffiltext{1}{Based on observations obtained at the Gemini Observatory, which is operated by the Association of Universities for Research in Astronomy, Inc., under a cooperative agreement with the NSF on behalf of the Gemini partnership: the National Science Foundation (United States), the Particle Physics and Astronomy Research Council (United Kingdom), the National Research Council (Canada), CONICYT (Chile), the Australian Research Council (Australia), CNPq (Brazil), and CONICET (Argentina).}



\begin{abstract}

The majority of atomic hydrogen Br$\gamma$ emission detected in the spectra of young stellar objects (YSOs) is believed to arise from the recombination regions associated with the magnetospheric accretion of circumstellar disk material onto the forming star.  In this paper, we present the results of a K-band IFU spectroscopic study of Br$\gamma$ emission in eight young protostars:  CW Tau, DG Tau, Haro 6-10, HL Tau, HV Tau C, RW Aur, T Tau and XZ Tau.  We spatially resolve Br$\gamma$ emission structures in half of these young stars and find that most of the extended emission is consistent with the location and velocities of the known Herbig-Haro flows associated with these systems.  At some velocities through the Br$\gamma$ line profile, the spatially extended emission comprises 20\% or more of the integrated flux in that spectral channel.  However, the total spatially extended Br$\gamma$ is typically less than $\sim$10\% of the flux integrated over the full emission profile.  For DG Tau and Haro 6-10 S, we estimate the mass outflow rate using simple assumptions about the hydrogen emission region, and compare this to the derived mass accretion rate.  We detect extended Br$\gamma$ in the vicinity of the more obscured targets in our sample and conclude that spatially extended Br$\gamma$ emission may exist toward other stars, but unattenuated photospheric flux probably limits its detectability.

\end{abstract}


\keywords{stars: pre-main sequence --- stars: winds, outflows --- stars: formation --- stars: individual (CW Tau, DG Tau, Haro 6-10, HL Tau, HV Tau C, RW Aur, T Tau, XZ Tau)}


\section{Introduction}

H~{\scshape i} emission lines are one of the defining characteristics of the classification of pre-main-sequence Sun-like sources known at T Tauri stars (TTS).  Still in the midst of formation, the less evolved TTS, known as classical TTS (cTTS), are surrounded by optically thick disks of gas and dust.  In most cases, these young pre-main-sequence stars are still interacting with and accreting matter from the innermost regions of their disks via stellar magnetic fields.  In this magnetospheric accretion paradigm, the stellar magnetosphere guides disk material from the inner disk onto the stellar surface through magnetic channels.  The gas travels along these channels or so-called accretion columns near free-fall velocities, terminating in an accretion shock at the stellar surface.  It is generally accepted that the gas is heated and ionized prior to and after reaching the stellar surface and that the characteristic H~{\scshape i} emission lines result, in part, from recombining and accreting hydrogen gas confined to these magnetic channels \citep{lynd1974, uchi1984, bert1988}.

Balmer H$\alpha$ emission is the dominant H~{\scshape i} feature present in the optical spectra of cTTS, and the emission line strength (or line width) is often invoked as a measure of accretion rates for these sources \citep{muze1998a}.  Spatially resolved observations of H$\alpha$ emission lines in these young stellar objects (YSOs) also show it to be a strong component of the optical line emission from outflows, indicating that the underlying stimulation mechanisms for the H~{\scshape i} lines is likely to be a combination of phenomena.   Magnetospheric accretion models can successfully reproduce many aspects of the H~{\scshape i} emission features detected in the spectra of young stars.  However, they notably fail to account for the highest velocity gas in the H~{\scshape i} line wings \citep{muze1998a}.  Spectro-astrometric observations of the Pa$\beta$ emission feature in the cTTS DG~Tau show that the high velocity blue-shifted gas ($v$~$>$~-200~km~s$^{-1}$) is extended in the same direction as [Fe~{\scshape ii}] at 1.644~$\mu$m, a forbidden emission line feature that traces the known outflow in the DG~Tau system.  This illustrates why a model producing H~{\scshape i} emission from accretion funnels does not account for high velocity gas forming the H~{\scshape i} line wings \citep{whel2004}.  While the high velocity gas is spatially shifted by as much as 0.$"$5 around DG Tau (or $\sim$~70~AU at the distance of Taurus-Auriga), the majority of the Pa$\beta$ emission remains coincident with the source.  Although these results demonstrate that there are multiple processes for stimulating H~{\scshape i} emission lines, the bulk of the emitting gas does seem to arise from radii within 14~AU of the central source constraining the emission to the magnetospheric accretion columns, inner gaseous disk, and the base of disk winds and outflows.

In the infrared, Br$\gamma$ (2.16~$\mu$m) emission serves as a surrogate for H$\alpha$ as a signpost for circumstellar disk accretion in TTS \citep{naji1996}.  Br$\gamma$ line luminosities appear to correlate with mass accretion luminosity in brown dwarfs, cTTS, and Herbig Ae/Be stars \citep[HAEBEs;][]{muze1998a, natt2004, moha2003, moha2005}.  Moreover, Br$\gamma$ and other lines in the infrared are less affected by optical depth effects that have proven problematic for using Balmer series lines to infer temperatures, densities, and geometries of the emitting gas.   Historically, the problem of predicting the spectra emerging from a recombining hydrogen gas has been divided in to two distinct cases, A and B \citep{bake1938}.  Case A theory applies to very low density gases that are optically thin to all transitions of the hydrogen atom, including the ultraviolet photons associated with the Lyman series transitions.  Case B theory, which applies to higher density gases which are optically thick to UV photons but optically thin to all n $\ge$ 2 transitions, is often applied to the environments of T Tauri stars.   In a Case B model of a recombining atomic hydrogen gas, Br$\gamma$ is $\sim$0.8-1\% of the flux of H$\alpha$ for a wide range of densities and temperatures (10$^2$ $<$ n$_e$ $<$ 10$^6$~cm$^{-3}$, 5000K $<$ $T$ $<$ 20000K).  H$\alpha$ is known to be a strong component in optical line emission from YSO outflows \citep{naji1996}, it seems natural that a corresponding component of the Br$\gamma$ emission would also arise from the outflows.  Yet, to date, there has been little evidence in the literature for spatially resolved Br$\gamma$ emission in the vicinity of YSOs, and hence nearly all Br$\gamma$ emission is assumed to arise from magnetospheric processes within the inner accretion zone.  In fact, infrared interferometric observations of HAeBe stars reveal that the Br$\gamma$ often arises from very compact locations within the dust sublimation radius of the circumstellar disk \citep{eisn2009, krau2008}.   Though, a small extended Br$\gamma$ emission component can not be ruled out based on these observations.  Further interferometric and spectro-astrometric programs that seek to reveal the inner $\sim$1 AU environments show that the compact Br$\gamma$ is not always well modeled by disk emission alone \citep{eisn2010, malb2010}.  The analysis implies that a non-negligible component from outflowing gas needs to be incorporated into the models to explain the Br$\gamma$ emission structure.

As an extension of the spectro-astrometric and interferometric studies mentioned above, imaging spectroscopy of these TTS can provide us with considerable insight into the geometric distribution of  H~{\scshape i} emitting gas in accreting systems.  Three-Dimensional imaging spectroscopy techniques can help to disentangle the respective contributions to the H~{\scshape i} emission features.  With just one pointing of a telescope, imaging spectroscopy with integral field units (IFUs) can provide three-dimensional x, y, $\lambda$ datacubes at high spatial resolution with simultaneous coverage of many emission lines of interest. There has recently been an increase in the capabilities for adaptive optics (AO) fed near IR integral field spectroscopy at 8-10 meter class observatories \citep{eise2000, mcgr2003, lark2006}.  IFUs optimized for AO spectroscopy have the power to spatially resolve emission line structures with less than 0.$''$1 extents over the full wavelength ranges sampled by typical IR spectrographs.  As such, the new generation of IFUs provides the means to study the accretion and outflow environments in cTTSs.

In this paper, we present detections of spatially resolved Br$\gamma$ emission in YSO environments from data acquired using the Near IR Integral Field Spectrograph at the Gemini North Observatory.  We report on Br$\gamma$ arising from eight classical TTS systems, and particularly highlight the spatially extended emission detected in four of these:  DG Tau, Haro 6-10 (also known as GV Tau), HL Tau and HV Tau C. 

\section{Observations}

Observations of the eight CTTSs listed in Table~1 were obtained using the Near IR Integral Field Spectrograph (NIFS) at the Frederick C. Gillette Gemini North Telescope on Mauna Kea, Hawaii.   NIFS is an image slicing IFU fed by Gemini's Near IR adaptive optics system, Altair, that is used to obtain integral field spectroscopy at spatial resolutions of $\le$0.$''$1 with a spectral resolving power of R$\sim$5300 at 2.2~$\mu$m \citep[as measured from arc and sky lines;][]{mcgr2003}.  The NIFS field has a spatial extent of 3$''\times3''$, and the individual IFU pixels are 0.$''$1$\times$0.$''$04 on the sky.  Data were obtained at the standard K-band wavelength setting for a spectral range of 2.003-2.447~$\mu$m.  All observations were acquired in natural seeing of better than 0.$''$7 for excellent AO correction.

The data sets for this study were acquired for commissioning and system verification of NIFS in October 2005 and February 2006, GTO time in December 2006, and in queue mode in February 2007 (see Table~1).  For each observation, a standard set of calibrations were acquired using the Gemini facility calibration unit, GCAL.  The raw IFU frames were reduced into datacubes using the NIFS tasks in the Gemini IRAF package\footnote{Information on the Gemini IRAF package is available at http://www.gemini.edu/sciops/data/dataIRAFIndex.html}.  

\cite{beck2008} discuss these NIFS data on DG~Tau, HL~Tau, HV~Tau~C, RW~Aur, T~Tau and XZ~Tau in the context of spatially resolved molecular hydrogen emission lines.  Hence the observational details, calibration, and data reduction information is described in great detail in that paper and excluded here.  To study the Br$\gamma$ emission in the young stars, the absorption features from the A0 stellar type telluric calibration stars were removed by fitting and dividing Voigt absorption profiles in the 2.16~$\mu$m spectral region and cleaning the calibration spectra for any small residuals.  Observations of Haro~6-10 were acquired with the laser-fed AO system using the R$\sim$16.5 mag southern component in this 1$.''2$ binary as the laser tip-tilt reference star.  These data were obtained in excellent laser-quality weather, photometric with better than 0$.''5$ seeing.  The CW~Tau and Haro~6-10 data were processed in a similar manner to all other data, as described above and in \cite{beck2008}.  All or part of the data for each source in this project was observed during photometric conditions and were flux calibrated using K-band magnitudes estimated by comparison to the brightness of the A0 standard star used for telluric correction.     With the exception of Haro 6-10, the derived fluxes of the systems (combined in the case of multiples) were within 10-15\% of published or 2MASS magnitudes.  For Haro~6-10, the NIFS IFU spectra were compared to infrared K-band images acquired nearby in time for a complimentary project, the difference in flux was less than 0.2 magnitudes between the two observations.  We estimate that our overall data flux calibration is good to $\pm$10-15\%. The data cubes were interpolated onto a square pixel grid with 0.$''$05 spatial sampling, and the velocity channel steps through the IFU cubes are $\sim$29~km~s$^{-1}$ pixels in extent at 2.20~$\mu$m ($\sim$56~km~s$^{-1}$ 2 pixel resolution).  The final reduced, combined, telluric corrected, and flux-calibrated datacubes of the Br$\gamma$ line emission for each target are discussed in detail in the following sections.
 
%% In this section, we use  the \subsection command to set off
%% a subsection.  \footnote is used to insert a footnote to the text.


%% This section also includes several of the displayed math environments
%% mentioned in the Author Guide.

\section{Spatially Extended Br$\gamma$ Emission from YSO Environments}

The data acquired for this project, described in detail in the preceding section, were obtained with the goal of studying the K-band features of molecular hydrogen emission.  The IFU spectral data were discussed by \citet{beck2008} in this context.  We never expected to detect spatially extended Br$\gamma$ emission in any CTTS.  This was a serendipitous discovery, revealed as we stepped through the raw velocity cube at Br$\gamma$ wavelengths in the Haro~6-10~S data.  The clear detection of spatially extended emission from the Haro~6-10~S jet prompted us to take a closer look at the Br$\gamma$ emission in all TTS for which K-band IFU spectra had been obtained.  As presented and discussed in the following, we have found significant spatially extended Br$\gamma$ emission in four of the eight stars presented here:  DG~Tau, Haro~6-10, HL~Tau and HV~Tau~C.  We do not find appreciable spatially extended Br$\gamma$ in CW~Tau, T~Tau, XZ~Tau, or RW~Aur.

Figures~1 through 4 show the images of the spatially extended Br$\gamma$ emission from DG~Tau (Figure~1), Haro~6-10 (Figure~2), HL~Tau (Figure~3) and HV~Tau~C (Figure~4).  The panels in these figures show:  a) - the continuum emission with contours of [Fe~{\scshape ii}] emission overplotted to demonstrate the outflow position and geometry, b) - continuum subtracted Br$\gamma$ line emission maps with contours of the continuum overplotted, c) -  ``point-source subtracted'' spatially extended maps of Br$\gamma$ emission with the contours of the continuum-subtracted Br$\gamma$ overplotted, and d) - images of the integrated blue-shifted Br$\gamma$ emission only, with contours of the point-source subtracted Br$\gamma$ emission overplotted.  At the right side of each image is the key correlating the image intensity to flux.   The continuum images in panels (a) were constructed by fitting a straight line to the continuum around the Br$\gamma$ emission feature and integrating the linear fit through the velocity channels that correspond to the Br$\gamma$ emission.  The images of the total Br$\gamma$ emission presented in the b) panels were constructed by subtracting the linear fit to the continuum from the datacubes, then integrating in velocity over the Br$\gamma$ emission feature.  The "point-source subtracted" Br$\gamma$ images shown in the c) panels were derived by normalizing the continuum (PSF) image to the peak flux in each velocity channel through the Br$\gamma$ emission datacube, subtracting this scaled continuum image off of the Br$\gamma$ cube, and integrating over the velocity extent to form an image of only the extended emission.  The image of spatially extended blue-shifted Br$\gamma$ emission shown in panel d) was made by integrating the 'point-source subtracted' image in 2-3 velocity channels of blue-shifted emission only.  The d) panels show that the spatially extended Br $\gamma$ emission is stronger in the blue-shifted velocity channels.  Because of Poisson statistics associated with the subtraction process, detection of spatially extended Br$\gamma$ emission is less robust at distances of $<$0.$"$1 from the central point-source.  The Br$\gamma$ emission in the vicinity of HV~Tau~C is quite weak, and all of the detected line emission is spatially extended.  As a result, Figure~4 presents only the (a) and (b) panels for HV~Tau~C.

For DG~Tau and Haro~6-10 S, the majority of the Br$\gamma$ emission that we detect, integrated over wavelength, is consistent with the stellar point-source image.  For HL Tau and HV Tau C, the majority of the Br$\gamma$ is not coincident with the central TTS.   The integrated Br$\gamma$ emission from DG Tau follows the point-source continuum contours with little deviation (Figure~1b). The spatially extended Br$\gamma$ emission from DG~Tau is detected in panel 1(c), and seen clearly in the blue-shifted emission shown in panel 1(d).  The Blue-shifted emission (Figure~1d) extends to the south-west of DG~Tau, at an orientation and velocity consistent with the known, collimated blue-shifted jet.  Haro~6-10 also shows the bulk of the Br$\gamma$ emission arising from the two stellar point-sources.  However, strong emission comes from the same location as the [Fe~{\scshape ii}] outflow emission from Haro~6-10~S (Figure~2c).  In fact, for Haro~6-10~S, the Br$\gamma$ arising from the outflow is $\sim$20\% of the total spatially integrated Br$\gamma$ emission in some blue-shifted velocity channels (Figure~2d). No spatially extended Br$\gamma$ is detected toward the Haro~6-10~N component.  

Data on HL~Tau were acquired with the 0.$"$2 occulting disk blocking the central stellar point-source, however an appreciable amount of the detected Br$\gamma$ emission is seen to deviate from the central stellar position.  HL~Tau shows strong spatially extended Br$\gamma$ that follows the scattered light nebulosity revealed in the continuum emission \citep[Figure~3b, 3c;][]{taka2007, clos1997}.  The integrated blue-shifted Br$\gamma$ from HL~Tau (Figure~2d) shows very weak emission detected at a 4$\sigma$ level of significance that corresponds precisely with the spatial location of the [Fe~{\scshape ii}] emission.  HV~Tau~C is a system with a known circumstellar disk viewed nearly edge-on \citep{stap2003} and the stellar continuum flux is much fainter than the other sources.  Curiously, we find no significant Br$\gamma$ emission associated with the locations of the continuum flux.  The Br$\gamma$ emission from HV~Tau~C is quite weak.  It is spatially extended from the continuum and it appears to only follow the location of the outflow seen in [Fe~{\scshape ii}] (Figure~4). 

The spatially extended Br$\gamma$ emission revealed in Figures~1, 2, and 4 for DG~Tau, Haro~6-10~S, and HV~Tau~C, lies precisely along the outflow axis of the known Herbig-Haro energy flows associated with these young stars.  For HL~Tau, the detected (low signal-to-noise) Br$\gamma$ emission in the blue-shifted component of the emission (Figure~3d) arises from the same spatial location as the known blue-shifted outflow \citep{taka2007}.  However in HL~Tau the majority of the spatially extended Br$\gamma$ emission (Figure~3c) is not appreciably shifted in velocity from the nominal stellar radial velocity; it is detected at much higher S/N, and appears to arise from Br$\gamma$ emission from the central point-source that has been scattered off of the wall of the outflow cavity \citep{clos1997, taka2007}.

Figure~5 shows the velocity profiles of the Br$\gamma$ flux associated with the central point source for DG~Tau, Haro~6-10, HL~Tau and HV~Tau~C at the location of the peak continuum emission (upper panel) and the profiles of Br$\gamma$ emission extracted in 0.$"$2 diameter apertures at ``Position B" and ``Position C" as designated in Figures~1 through 4 for each star.  The spatially extended Br$\gamma$ from DG~Tau and Haro~6-10~S is blue-shifted in velocity by $>$~100~km~s$^{-1}$ with respect to the central point-source flux.  HL~Tau shows strong emission at the velocity of the point-source flux in ``Position B," and very slight $\sim$3-4$\sigma$ detection of flux from blue-shifted ($\sim$-200~km~s$^{-1}$) emission from the outflow. 

In DG~Tau, Haro~6-10~S and HL~Tau, the spatially extended Br$\gamma$ emission that we detect from the outflows corresponds to the blue-shifted regions of the jets.  This is consistent with the fact that blue-shifted jet components are flowing into our line of sight, and are hence less obscured by intervening circumstellar disk material.  HV~Tau~C is the only source where both blue and red-shifted Br$\gamma$ is detected from opposite sides of the outflow with respect to the (estimated) position of the central star.  The extended Br$\gamma$ from HV Tau C is quite weak in some regions, it is detected at a low signal-to-noise but follows the location of the extended outflow.  Curiously, we find that the blue-shifted outflow lies to the northeast, and the red-shifted emission is to the southwest, which \citep[as predicted by][]{stap2003} is at odds with the relative brightnesses of the lobes of the scattered light edge-on disk reflection nebula associated with this source.  The southwestern lobe of the HV Tau C scattered light nebula is brighter, and was thus thought to be associated with the blue-shifted (closer) lobe of the outflow.  Our data reveal that this is not the case, and the north-eastern, fainter lobe seems to be associated with more blue-shifted outflow emission.

Figure~6 presents the standard deviation of continuum subtracted Br$\gamma$ flux with increasing distance from the central star (plotted as a solid line) for HL~Tau (a),  DG~Tau (b), Haro~6-10 (c) and HV~Tau C (d).  These curves were derived for each source by calculating the mean flux within 0.$"$08 wide pixel annuli from the central stellar position and computing the standard deviation from the mean.  This standard deviation of continuum subtracted line emission provides a measure of the uncertainty in the subtraction process, and thus can be used to gauge the S/N of the spatially extended line emission.  Though, the continuum subtracted plots do have the extended Br$\gamma$ flux included within the field.  As a result, the measured standard deviations in the spatial bins with appreciable Br$\gamma$ flux are correspondingly higher, and the accuracy of the subtraction process is thus under-estimated. Overplotted in each of the four panels is a dashed line that shows the magnitude of the point-source subtracted Br$\gamma$ flux with increasing distance from the central stellar position.  Comparison of the dashed curve with the solid curve provides an estimate of the measured S/N of the extended Br$\gamma$ emission.  The peak S/N for extended Br$\gamma$ is $\sim$7 for HL~Tau, $\sim$5 for DG~Tau, $\sim$25 for Haro~6-10 S and $\sim$4 for HV~Tau~C.

Figure~7 plots the same standard deviation of continuum subtracted Br$\gamma$ flux with increasing distance from CW~Tau, T~Tau, XZ~Tau, and RW~Aur, and a dashed line is also over-plotted that presents the peak magnitude of the point-source subtracted Br$\gamma$ emission from the central star (in the case of the multiple systems, the central star is assumed to be the brightest stellar component).  The stellar companion position locations are apparent in the plots for XZ~Tau and T~Tau, while RW~Aur's companion has a slightly greater separation than presented in Figure~7d.   For all four sources presented in Figure~7, no excess was seen in spatially extended Br$\gamma$ emission at the location of the known Herbig-Haro outflows.  For the cases of CW~Tau, XZ~Tau and RW~Aur, no strong evidence of extended Br$\gamma$ emission beyond a S/N of $\sim$3 is found.   Curiously, XZ~Tau~B showed weak Br$\gamma$ emission associated with the position of the star, but XZ~Tau~A had no detectable Br$\gamma$.  This result is seemingly at odds with the proposition that XZ~Tau~A is the more actively accreting star and the main driving source of the Herbig-Haro flow associated with this system \citep{kris2008}  Although RW~Aur exhibits very strong, centralized Br$\gamma$ emission flux, RW~Aur~B showed no measurable Br$\gamma$ emission associated with stellar mass accretion from its circumstellar disk.

The central region of Br$\gamma$ emission for T~Tau~N was saturated in the data, making a proper measurement of the continuum subtracted Br$\gamma$ flux and point-source subtracted flux difficult.  The surrounding spatial and spectral regions nearby are not saturated, and were used to estimate the 2.16~$\mu$m flux level based on the shape of the PSF.  A cursory fitting analysis done at velocities on and off of the emission line using the A 0 spectral type telluric calibrator as a PSF reference showed that the Br$\gamma$ emission associated with T~Tau~N is not appreciably extended compared to the point-source continuum emission.   The line emission is asymmetric around the T~Tau~S PSF, and is brighter to the northwest at the location of T~Tau~Sb.  Hence, the Br$\gamma$ line emission associated with the nearby 0$."$1 T~Tau~S a+b binary seems to arise preferentially from the $\it{b}$ component, and it is stronger than the continuum flux (i.e., the T Tau Sb / Sa flux ratio is greater in Br$\gamma$).  This causes the apparent enhancement in Br$\gamma$ emission at the position of the companion in Figure~7b.  Extraction of the individual spectra of the blended components was not done because of the saturation of T~Tau~N, which would need to serve as a PSF calibrator.

\section{Br$\gamma$ Estimates of YSO Mass Accretion and Mass Outflow Rates}

\citet{muze1998b} showed that the Br$\gamma$ line luminosity from young stars correlates with the mass accretion rate, as determined from UV excess emission.  Hence, we now use our detected Br$\gamma$ line fluxes to derive mass accretion rate for the observed targets.  Table~2 presents the total Br$\gamma$ emission line fluxes (i.e., integrated over all velocity channels) for each of the young stars in this study.  The relation from \citet{muze1998b} assumed all detected Br$\gamma$ emission was associated with the stellar point-sources, so the line fluxes that we have included in Table~2 are the total integrated flux values, including emission detected in the outflowing gas.  Also included in Table~2 are the adopted stellar parameters used to derive the mass accretion rates:  the stellar mass, temperature, luminosity and visual extinction \citep{keny1995, whit2001, hart2003, dopp2008}. The stellar parameters seem consistent with our K-band spectra, so we do not rederive them from our data.

XZ~Tau~B, T~Tau~S and Haro~6-10~N are ``infrared luminous companions'' (IRCs) to their respective primaries, and the stellar parameters for these sources are much less certain \citep{kore1997,whit2001}.  The accretion activity and line of sight visual extinction toward the IRCs could also be variable \citep{ghez1991, beck2001, lein2001, beck2004}.  Moreover, T~Tau~S is itself a binary, and the bulk of the Br$\gamma$ emission that we detect arises not from the IRC but from the $\sim$M-type T~Tau~Sb companion \citep{duch2005}.  Haro~6-10~N was found by \citet{dopp2008} to have weak evidence of photospheric Na absorption features at 2.20~$\mu$m, with a late spectral type and an infrared veiling value estimated to be in the range of 12-15.  Similarly, XZ~Tau~B also has strong infrared veiling and a poorly determined spectral type.   As a result, we do not try to estimate mass accretion rates for T~Tau~S, Haro~6-10~N or XZ~Tau~B.  Column~7 of Table~2 presents the mass accretion rates ($\dot{M}_{acc}$) derived for all of the other stars, using the relation from \citet{muze1998b} for the Br$\gamma$ line luminosity to accretion luminosity and the Virial Theorem treatment put forth by \citet{gull1998}.

The mass accretion rates derived for the eight stars in this study that have well determined stellar parameters lie in the range from less than 4$\times$10$^{-10}$~M$_{\odot}$ for RW~Aur~B, to 1.5$\times$10$^{-7}$~M$_{\odot}$ for T~Tau~N, with most sources in the range of 10$^{-8}$ to 10$^{-7}$~M$_{\odot}$.  These values compare lie within the overall range of mass accretion rates derived for cTTSs \citep{muze1998b, gull1998}.  For the most part, the mass accretion rates that we derive here are similar to $\dot{M}_{acc}$ values for these stars that have been derived from previous studies, within the associated uncertainties (e.g., Hartigan et al.\ 1995; Muzerolle et al.\ 1998; White et al.\ 2004).   However, direct comparisons between mass accretion values for DG~Tau, particularly, show a large discrepancy in our study.  DG~Tau has in the past exhibited a mass accretion value on the high side of the range for TTSs;  around or just under 10$^{-6}$~M$_{\odot}$~yr$^{-1}$.  Our derived mass accretion rate is an order of magnitude less than many previously published estimates \citep{hart1995,muze1998b}.  DG~Tau is known to vary in flux and spectral characteristics on short time scales \citep{bisc1997, hess1997, bary2008}.  Bary et al.\ (2008) found a nightly average Br$\gamma$ equivalent width of 6.4 \AA\ for seven nights of observations with a standard deviation of 20\%.  Therefore, the discrepancy we see in mass accretion rate toward DG Tau might be a result of both intrinsic stellar and accretion variability. 

Also presented in column~8 of Table~2 is the Br$\gamma$ flux from the spatially extended outflows from DG~Tau and Haro~6-10~S.  For our analysis of Br$\gamma$ emission seen in the outflows, we extracted a region of Br$\gamma$ flux from the spatially extended jets from DG~Tau and Haro~6-10 (Figures~1d and 2d).  Multi-epoch observations have shown that the average transverse proper motion of an outflow is $\sim$0$."$2 to 0$."$3 per year, or 28-42~AU at the $\sim$140pc distance of Taurus \citep{torr2009}.   Both the DG~Tau and Haro~6-10~S outflows are viewed at 60-70$^{\circ}$ inclination angles \citep{movs1999, pyo2003}, and the 28-42 AU extent in the outflow is also roughly consistent with the corresponding annual motion of an average flow velocity of $\sim$200~km~s$^{-1}$; e.g., material in the flow would move 28-42AU/yr at the estimated geometry and velocity.   For this reason, to capture the emitting flux from approximately one year of jet motion, we have extracted a segment, 0$."$3 in extent, of the spatially resolved Br$\gamma$ emission along the jet axes for the DG~Tau and Haro~6-10 outflows.  The width of the extraction box corresponds to a 0$."$2 radius encircling the jet.  The Br$\gamma$ emitting volume, V$_{\rm {HI}}$, roughly corresponds to this area surrounding the jet, and is approximated by a cylinder of 28~AU (0$."$2) radius and 42~AU (0$."$3) height.  The ``one year" integrated Br$\gamma$ outflow emission fluxes for DG~Tau and Haro~6-10~S extracted from these apertures are 6.8$\times$10$^{-16}$ and 8.5$\times$10$^{-16}$~erg~s$^{-1}$~cm$^{-2}$, respectively.  

Under the assumption that the spatially extended Br$\gamma$ emission seen from DG~Tau and Haro~6-10~S arises from thermally excited emission from optically-thin, post-shock regions of the outflows and behave as a Case~B recombining gas \citep{bake1938, broc1971, humm1987, stor1995}, we use a simple analysis to measure the mass outflow rate from these stars.  Using the H~{\scshape i} emission coefficients from \cite{oste1989}, the expression for the detected Br$\gamma$ flux can be described as:

\begin{equation}
\rm F_{\rm Br\gamma}=1.2\times10^{-28}(\rm N_e^2V_{\rm {HI}}/D^2)
\end{equation}

\noindent
where D is the distance to the emitting region in centimeters, N$_e$ is the electron density, and V$_{\rm HI}$ is the volume of the emitting region (MKS units).  The electron temperature is assumed to be $\sim$10$^4$~K, which is a reasonable estimate for the inner regions of YSO outflows.  The mass of the emitting hydrogen in the spatially resolved regions of the outflow can be estimated as M$_{\rm HI}$~=~$m_p$N$_e$V$_{\rm HI}$.  Merging this equation for the mass with the above equation for the flux gives:

\begin{equation}
 M_{\rm {HI}} = 1.5\times10^{-13} (\rm F_{\rm Br\gamma}V_{\rm HI})^{1/2}D
\end{equation}

\noindent
We can thus solve for the mass of emitting hydrogen gas using the parameters of our measured Br$\gamma$ flux, the selected emission volume, and the assumed 140~pc distance to the stars, which is based on distances derived toward TTS within the Taurus star forming complex \citep{torr2009}.   For DG~Tau, an H~{\scshape i} mass of 1.2$\times$10$^{-8}$~M$_{\odot}$ is derived, and for Haro~6-10~S this value is 1.4$\times$10$^{-8}$~M$_{\odot}$.  The flux extraction volumes were chosen to be the same, so the difference in the atomic hydrogen outflow masses between the two stars is determined by the difference in their detected Br$\gamma$ flux values.  Having chosen the flux extraction volumes to correspond to the average $\it{annual}$ proper motions of these jets, these estimates approximate the mass outflow rates for DG~Tau and Haro~6-10~S in solar masses per year.

Based on this analysis, we derive average electron densities on the order of a few $\times$10$^4$~cm$^{-3}$ for the DG~Tau and Haro~6-10~S outflows.  These densities are consistent with the large values of 10$^4$ to 10$^6$~cm$^{-3}$ for N$_e$ that are often found within the inner one hundred AU regions of young star outflows derived by inspecting [Fe II] and other forbidden emission species \citep{hart1995, bacc1999, hart2007, hart2009}.   These electron densities are also very consistent with past values found in the inner DG~Tau high velocity blue-shifted jet \citep{bacc1999, coff2008}.  

The hydrogen outflow rates of 1.2$\times$10$^{-8}$~M$_{\odot}$~yr$^{-1}$ for DG~Tau and 1.4$\times$10$^{-8}$~M$_{\odot}$~yr$^{-1}$ for Haro~6-10~S represent lower limits for the true mass outflow levels derived in this manner.   These mass outflow rates are underestimates of the true mass flow because only the fraction of the gas that has recently been heated by the shock radiates in the emission species that is studied.  We also only detect the gas in the high velocity component of the outflow, not in lower velocity flow surrounding the jet axis \citep{bacc2000, pyo2003}.  Moreover, the derived Br$\gamma$ flux in the outflowing volume was not corrected for any line of sight extinction effects which might further serve to increase the derived mass outflow rate, particularly in the obscured Class~I star, Haro~6-10~S.  Thus, we predict that the true  $\dot{M}_{out}$ values are greater than 1.2$\times$10$^{-8}$M$_{\odot}$~yr$^{-1}$ for DG~Tau and more than 1.4$\times$10$^{-8}$M$_{\odot}$~yr$^{-1}$ for Haro~6-10~S.

\section{Discussion}

This study shows that not all H~{\scshape i} Br$\gamma$ emission from classical TTS arises from magnetospheric accretion processes within a few radii from the central star.  We detect spatially resolved Br$\gamma$ from DG~Tau, Haro~6-10~S, HL~Tau and HV~Tau~C, which represents 50\% of the TTS systems in our sample.  We do not spatially resolve Br$\gamma$ line emission in the environments of XZ~Tau, RW~Aur, T~Tau and CW~Tau.  Two stars within these latter systems, RW~Aur~B and XZ~Tau~A, exhibit no Br$\gamma$ emission at all. 

In two of the blue-shifted velocity channels, the spatially extended Br$\gamma$ from Haro~6-10~S makes up $\sim$20\% of the total spatially integrated line flux.  Integrated over the whole velocity width,  about 10$\pm$2\% of the Br$\gamma$ emission from Haro~6-10~S is spatially resolved at distances of greater than 0.$"$1 (14 AU) in the extended outflow (entirely from the blue-shifted velocity component).   The spatially extended Br$\gamma$ emission makes up $\sim$2\% of the total line flux for DG~Tau.  All of the detected Br$\gamma$ flux seems to arise from the outflow for HV~Tau~C, but the star is not seen directly because the continuum flux is measured only from the scattered light nebulosity.   The inner magnetospheric accretion region of HV~Tau~C may be shielded by the inner rim of the central circumstellar dust disk in our edge-on viewing orientation, hence in this case we might not see the strong central Br$\gamma$ component in the scattered light nebulosity.  HL~Tau has considerable Br$\gamma$ emission scattered off of its surrounding outflow cavity walls, and the fraction of spatially extended Br$\gamma$ emission is estimated to be 18$\pm$7\% of the line flux from the central position.  Observations of HL~Tau were acquired with an occulting disk in the beam, so the total point source flux is estimated from the PSF shape in the target acquisition setup images and the resulting value is significantly less certain.  

The majority of the integrated Br$\gamma$ flux that we measure is spatially unresolved from the position of the central stellar sources in our data, with the noted exception of the edge-on disk system HV~Tau~C.  NIFS data can spatially resolve the bright Br$\gamma$ emission beyond $\sim$0.$"$1 from the star, or about $\sim$14~AU.   Additional Br$\gamma$ emission likely arises from the outflows in regions closer than our resolution limit, where it cannot be detected.  Overall, the extended Br$\gamma$ line emission that we detect beyond $\sim$14~AU distances from the parent stars comprises anywhere from a few percent  (DG~Tau) to all of the detected line flux from these systems (HV~Tau~C).  

All of the sources that we have studied here are known to drive Herbig-Haro outflows.  It is not clear why some stars exhibit extended Br$\gamma$ emission, while other sources with strong and collimated outflows do not.  However, from information presented in Table~2, we see that the stars that have appreciable spatially extended Br$\gamma$ emission also have stronger estimated levels of optical visual extinction, A$_{\rm v}$, toward the stellar photosphere.  The stars where we do not find extended Br$\gamma$ emission all have lower estimated optical obscurations.  In the systems where extended Br$\gamma$ emission is seen, the stronger levels of stellar continuum flux attenuation from the high visual extinction may make the weak spatially extended emission easier to detect.  The stars that do not exhibit extended Br $\gamma$ emission have brighter stellar continuum flux, which may be a result of less obscuration by natal material because of a slightly older evolutionary state, or a perhaps a more inclined viewing geometry that directly reveals more of the central photosphere of the star.  Thus, spatially extended Br$\gamma$ emission may exist toward the other stars, but the bright continuum flux might prevent us from detecting it.  

Several recent spectro-astrometric and interferometric investigations of inner YSO disks have sought to spatially resolve the atomic hydrogen associated with the magnetospheric recombination regions within 1~AU from the target stars \citep{whel2004, eisn2009, krau2008}.  The spectro-astrometric study of \citet{whel2004} revealed that an appreciable amount of Pa$\beta$ emission from the blue and red-shifted velocity profiles arises from spatially extended distances from a TTS.  Three of their four targets exhibited spatially resolved Pa$\beta$ emission, two revealed evidence for spatially extended bi-polar red and blue-shifted components.   This is an important finding, since the broadened emission line wings seen in the H~{\scshape i} profiles from cTTS have never been well explained by magnetospheric accretion models.  Evidence from this Pa$\beta$ study and now our Br$\gamma$ project suggest that the H~{\scshape i} emission from TTS does have non-negligible components from the outflow on spatial scales that were unresolvable prior to the current generation of sensitive instrumentation on 8-10 meter class telescopes.

Thus far, spectro-interferometric studies of Br$\gamma$ that can spatially resolve the inner magnetospheric accretion region have largely concentrated on the brighter Herbig Ae/Be stars (HAEBEs; \cite{eisn2009, krau2008, malb2007}).  Interestingly, most interferometric data reveals that the Br$\gamma$ emission surrounding the HAEBEs arises from a centralized location that is spatially more compact than the infrared continuum emission, with perhaps some contribution from an outflowing wind \citep{malb2007}.  The compact H~{\scshape i} likely arises predominantly from the accretion-driven processes within the star-disk boundary; either from the central stellar mass accretion engine or from a stellar or disk wind \citep{krau2008, malb2007, eisn2010}.  Further work is warranted to better reveal the central emission location of H~{\scshape i} from TTSs, and the fraction of H~{\scshape i} emission that is truly spatially extended from the parent star in comparison to the higher mass HAEBEs.

Although we do detect spatially extended Br$\gamma$ emission from 50\% of our sample targets, the bulk of the line emission from most stars does arise within $\sim$14~AU from the central unresolved point sources.  While the spatially extended H~{\scshape i} seems to affect the wings of the velocity line profile shapes \citep{whel2004}, it is likely that the majority of the low velocity H~{\scshape i} is emitted via magnetospheric processes in the central disk-star accretion region.  \citet{bary2008} presented a large-scale multi-epoch near-infrared spectroscopic survey of 15 actively accreting TTS in Taurus-Auriga.  They compared the H~{\scshape i} line ratios of 16 Paschen and Brackett series emission features and found little variations in these ratios from epoch to epoch and from source to source.  For the first time, a statistically significant correlation was found between the observed H~{\scshape i} line ratios in TTS systems and those predicted by Case~B line recombination theory \citep{bake1938,humm1987,stor1995} for a tightly constrained gas temperature and electron density.  While the range of electron densities Bary et al.\ find for the emitting H~{\scshape i} gas agrees with that predicted by magnetospheric accretion models, the temperatures are substantially lower than expected for accreting gas\footnote{\citet{bary2008} found the most likely temperature and density ranges of $T$~$<$~2000~K and 10$^9$~cm$^{-3}$~$<$~$n_e$~$\le$~10$^{10}$~cm$^{-3}$.}.  These results suggest that the emission is from a recombining gas not in local thermodynamic equilibrium and is instead being stimulated by a non-thermal process.  In addition, these results may also suggest a combination of emission sources for the emitting H~{\scshape i} gas in the central regions and that the observed H~{\scshape i} line fluxes are a superposition of emitting gas with different physical conditions.  As such, one might not necessarily expect a linear correlation between detected Br$\gamma$ flux from the central star and the liberated accretion luminosity measured from the UV excess emission, as found by \citet{muze1998b}.  However, the correlation does exist and the mass accretion rates we derived from our measured Br$\gamma$ line fluxes are consistent with those found in past studies of TTS.  Taken in combination, these results further confirm the connection between accretion and outflow activity. 

Deriving accurate mass outflow values is notoriously difficult.  This is in part because the emission line regions in outflows measure only material in the post-shock area of a jet and do not sample outflow material ahead of the shock.   Shocks from outflows and their associated optical and infrared emission lines are obvious observational manifestations of mass outflow activity, but they are the result of fast material plowing into slower clumps of gas in the outflow path.  Hence, material can (and does) exist in the outflow without these detectable emission features if there is no slower material in the flow path to cause the shock.  Methods described in the literature for measuring mass outflow rates from classical TTSs rely largely on measuring the mass flux from forbidden emission line species and converting this to an outflow rate based on estimates of the gas ionization fraction \citep{hart1995, cabr2002, doug2002, bacc2002, coff2008, agra2009}.   The different observational techniques derive mass outflow to mass accretion ratios in the range of 1 to 10\%.  In some cases, multi-wavelength or emission line studies of the same system find $\dot{M}_{out}$/$\dot{M}_{acc}$ values that differ by factors of $\sim$10, even for the same system (e.g., DG~Tau, see below).  Measurement of mass outflow rates using atomic hydrogen emission species has been difficult in the past because of the strong optical depth effects in optical Balmer emission species.  The infrared Brackett features are less optically thick, and may serve as a more robust tracer of the atomic hydrogen gas mass in the young star outflows.  

In the previous section, we derived the first masses of ionized hydrogen using Br$\gamma$ emission for the DG~Tau and Haro~6-10~S outflows.  Comparison of the mass accretion rates with our derived atomic hydrogen mass outflow rates for DG~Tau and Haro~6-10~S (see Table~2) reveals that the outflowing H~{\scshape i} mass is estimated at $\sim$10-15\% of the mass accreting onto the central star.  Moreover, our derived mass outflow rates are also lower limits, implying that the actual ratio of  $\dot{M}_{out}$/$\dot{M}_{acc}$ may be larger than 10-15\% for these two sources.    These ratios are consistent with $\dot{M}_{out}$/$\dot{M}_{acc}$ values predicted by the theory of mass accretion efficiency and magnetocentrifugal launching mechanisms (i.e., disk winds and accretion-driven stellar winds) for young star outflows \citep{koen1991, matt2005, cran2008}, and are on the high side in comparison with ratios found from other observational investigations  \citep{edwa1987, hart1995, cabr2002, doug2002, bacc1999, coff2008, agra2009}.

It is interesting that we find $\dot{M}_{out}$/$\dot{M}_{acc}$ values for DG~Tau and Haro~6-10~S that seem in the high range compared to ratios derived from other observations.  Additionally, the $\dot{M}_{out}$ rates we find are lower limits, as discussed in $\S$5.  However, the uncertainties in the derivation of both the  $\dot{M}_{out}$ and $\dot{M}_{acc}$ values can be very large.  Our lower limit for the value of M$_{out}$ in DG~Tau and Haro~6-10~S is derived from straightforward assumptions on the physics of LTE atomic hydrogen in recombination regions.  As noted above, the hydrogen recombination emission area very close to the central star is possibly excited by non-thermal processes in the inner magnetospheric region \citep{bary2008}.  Hence, it seems feasible that these non-LTE conditions may extend into the inner regions of the outflow, making our H~{\scshape ii}-like analysis of hydrogen emitting mass correspondingly uncertain.  This is likely one of the largest sources of uncertainty in our analysis, but it is also difficult to characterize.  We also estimate an additional source of error in our selection of the emission volume that was chosen to represent one year worth of jet motion.  Moreover, the correlation of integrated Br$\gamma$ line luminosity with stellar mass accretion luminosity has an intrinsic scatter \citep{muze1998b}, and deviation in $\dot{M}_{out}$/$\dot{M}_{acc}$ by a factor of several for a given target can result from using this method to derive the stellar mass accretion rate.  Multiple observations of accretion indicators from large sample of TTS demonstrate the highly variable nature of accretion activity in these young stars \citep[e.g.,][]{bary2008,nguy2009}.  However, the effects from intrinsic time variation in the accretion and outflow properties are not an issue in our study because the mass outflow and mass accretion rates are derived simultaneously from the same data set.

The blue-shifted jet emerging from DG~Tau is arguably one of the best studied outflows associated with a young star.   It was among the first cTTSs for which a collimated jet-like outflow was discovered \citep{mund1983}.  Since its discovery, the HH~158 outflow from DG~Tau has been investigated with high spatial resolution imaging and spectroscopy with $\it{HST}$, ground-based adaptive optics systems, and spectro-imaging techniques \citep{lava1997, lava2000, bacc2000, doug2002, bacc2003, pyo2003, taka2004, coff2008}.  This outflow has observationally revealed the structure of collimated YSO jets in great spatial detail;  they are typically comprised of an on-axis high velocity component (HVC) at radial velocities greater than $\sim$50~km~s$^{-1}$ that can extend to spatial distances of hundreds of AU from the star, and an encompassing shell of lower velocity gas that extends only about $\sim$100~AU away from the central star \citep{bacc2000, pyo2003, taka2004}.   Although He~I 10830 emission is commonly attributed only to stellar winds in the inner $\sim$10 AU environments of CTTS, \citet{taka2002} detect He~I 10830 emission from DG Tau at the jet velocity, spatially extended over ~ 0.5 arcsec  ($\sim$70AU) of the jet, though their spectroastrometry did not find extension in the low-velocity He I 10830 emission.  \citet{bacc2000} and \citet{pyo2003} described the blue-shifted HH~158 jet as having an ``onion-skin" structure, with high velocity low density gas on-axis, surrounded by successive layers of lower velocity, higher density gas.  Integrated over the width of the jet axis, the $\dot{M}_{out}$/$\dot{M}_{acc}$ values derived for the DG~Tau jet from past studies lie in the range of 0.05-0.1 \citep{lava1997, bacc2000, coff2008}, though most of these estimates use large mass accretion rate of $\sim$10$^6$ M$_{\odot}$~yr$^{-1}$ \citep{hart1995}.  \citet{bacc2003} analyzed the kinematics of DG~Tau's jet and found evidence for velocity asymmetries across the jet axis, and they interpret this result as evidence for rotation within the inner jet channel.  This controversial finding challenges the theories of magnetospheric interaction at the inner star + disk + outflow boundary region.     

The outflow from DG~Tau has served as a laboratory to test the physics of YSO jets, so it is fitting that our first discovery of spatially extended Br$\gamma$ from cTTSs includes this system.   In DG~Tau, the spatially extended Br$\gamma$ emission that we detect arises from the high velocity, on-axis gas in the blue-shifted outflow.  Interestingly, this H~{\scshape i} coincides very closely in projected location with the recently discovered X-ray emission in the jet \citep{gude2005, gude2008}.   \citet{gude2008} found that the  blue-shifted outflow from DG~Tau has emission from soft X-rays out to spatially extended distances of several hundred AU from the central star.  The temperatures of the soft X-ray emission regions associated with the extended X-ray jet are on the order of a few times 10$^6$K.  These temperatures pose a problem for understanding jet heating mechanisms.  The standard jet shock models for YSO outflows cannot explain temperatures on the order of 10$^6$K, and this argues strongly in favor of some form of non-thermal excitation of the gas; for example by Ohmic heating from magnetic dissipative currents \citep{gude2008}.   The detected Br$\gamma$ emission from the DG~Tau blue-shifted outflow seems to correspond to gas nearby to these super-heated regions, and thus if strong magnetic fields significantly effect the physical conditions in the inner flows then our simple method of deriving $\dot{M}_{out}$ assuming thermal Case~B recombination of atomic hydrogen at temperatures of $\sim$10$^4$K would be correspondingly inaccurate.  Further tests and comparison of the physical conditions in Br$\gamma$ and X-ray emitting jets could help clarify the physics in the inner regions of the outflows.  Unfortunately, as previously mentioned, HL~Tau, HV~Tau~C and Haro~6-10~S are Class~I protostars or cTTSs with circumstellar disks viewed in a nearly edge-on configuration, and all three of these systems have very strong optical obscuration (see Table~2).  It is likely not feasible to detect the soft X-ray emission from these jets to further test the physical conditions of the Br$\gamma$ emission environments.

\section{Summary}

We have presented the results of our K-band integral field spectroscopy study of the Br$\gamma$ line emission from eight T Tauri star systems.  The key points of our study are:

1) Using adaptive optics fed integral field spectroscopy, we have spatially resolved Br$\gamma$ line emission in the circumstellar environments around four of our eight survey target stars:  DG~Tau, Haro~6-10~S, HL~Tau and HV~Tau~C.

2)  The spatially extended Br$\gamma$ emission arises predominantly from the hydrogen recombination regions associated with the inner Herbig-Haro outflows from these young stars.  Only HL~Tau shows a significant contribution of stellar Br$\gamma$ emission scattered into our line of sight off of the outflow cavity walls.  This emission from HL~Tau has a similar morphology to the scattered light continuum flux, and it likely originated from the inner magnetospheric accretion region around the star.

3)  At some blue-shifted velocities, the spatially extended Br$\gamma$ emission comprises $\sim$20\% of the detected Br$\gamma$ (e.g., Haro~6-10).  Although we spatially resolve Br$\gamma$ emission from outflows in our high-contrast measurements, the majority of the integrated Br$\gamma$ from most systems is spatially unresolved and may arise from the magnetospheric accretion processes at the location of the central stellar source (with the exception of HV~Tau~C).

4)  All of the Br$\gamma$ emission that we detect above the continuum flux from HV~Tau~C is from the spatially extended emission, consistent with the location of the known Herbig-Haro outflow.  HV~Tau~C is seen in continuum light as a scattered light edge-on disk nebulosity.  The inner magnetospheric component of Br$\gamma$ flux may be shielded from our line of sight by material in the inner edge of the circumstellar disk.

5)  Derivation of the stellar mass accretion rates from the relationship between Br$\gamma$ line luminosity and mass accretion reveal $\dot{M}_{acc}$ values that are typical of cTTSs.

6)  Detection of the spatially extended Br$\gamma$ emission from the outflows in the DG~Tau and Haro~6-10~S systems have allowed us to derive a value for the emitting hydrogen mass outflow rate using simple arguments applicable to hydrogen recombination regions.  The corresponding values for $\dot{M}_{out}$/$\dot{M}_{acc}$ that we derive are on the order of $\sim$10-15\%, consistent with many prediction from accretion-driven stellar winds and disk winds, while on the high side in comparison to observationally derived mass outflow to accretion rate ratios from past studies.

7)  We find that in some young protostars, Br$\gamma$ emission extended on spatial scales of greater than 0.$"$1 (14 AU) can contribute $\sim$ 10\% of the flux to the detected integrated line emission (or more, as in the case of HL Tau and HV~Tau~C).


%% Included in this acknowledgments section are examples of the
%% AASTeX hypertext markup commands. Use \url without the optional [HREF]
%% argument when you want to print the url directly in the text. Otherwise,
%% use either \url or \anchor, with the HREF as the first argument and the
%% text to be printed in the second.

\acknowledgments

Several NIFS datasets presented in this study were acquired during the early stages of instrument integration at Gemini North Observatory, and we are extremely grateful for the support of the NIFS teams at the Australian National University, Auspace, and Gemini Observatory for their tireless efforts during the instrument commissioning and system verification.  This study is based on data from the Gemini Observatory, which is operated by the Association of Universities for Research in Astronomy, Inc., on behalf of the international Gemini partnership of Argentina, Australia, Brazil, Canada, Chile, the United Kingdom, and the United States of America.

\clearpage

\clearpage

\begin{deluxetable}{lcccccccc}
\tabletypesize{\scriptsize}
\tablecaption{Observing Log \label{tbl-1}}
\tablewidth{0pt}
\tablehead{
\colhead{Star} & \colhead{HH \#} & \colhead{Obs. Date} & \colhead{PA of} & 
\colhead {Exp Time/Coadds} & \colhead{\# Exp} & \colhead{Total Exposure} & 
\colhead{Note on} \\
\colhead{Name} & \colhead{} & \colhead{(UT)}  & \colhead{Obs.}   & 
\colhead {} & \colhead{} & \colhead{Time} & \colhead{Obs.} } 
\startdata
T Tau & HH 255 & 2005 Oct 25 & 0$^{\circ}$ &  5.3s/24 & 36 & 4580s & AO Flexure Test \\
RW Aur  & HH 229 & 2005 Oct 22 &   221$^{\circ}$ & 40s/1 & 11 & 440s & AO Guide Test \\
XZ Tau & HH 152 & 2005 Oct 25 &   0$^{\circ}$ & 30s/1 & 28 & 820s & AO+OIWFS$^*$ Guide Test \\
DG Tau & HH 158 & 2005 Oct 26 &   0$^{\circ}$ & 20s/6 & 101 & 12120s & AO+OIWFS$^*$ Flexure Test \\ 
HV Tau C & HH 233 & 2005 Oct 22  &  114$^{\circ}$ & 900s/1 & 3 & 2700s & System Sensitivity Test \\
HL Tau & HH 150 & 2006 Feb 12 &   146$^{\circ}$ & 900s/1 & 3 & 2700s & $0.''2$ Occulting Disk SV \\
CW Tau & HH 150 & 2006 Dec 5 &   146$^{\circ}$ & 900s/1 & 3 & 2700s & NIFS PI GT \\
Haro 6-10 & HH 150 & 2007 Feb 7 \& 8 &  0$^{\circ}$ & 300s/1 & 9 & 2700s & NIFS + LGS AO Queue \\
\enddata

%% Text for table notes should follow after the \enddata but before
%% the \end{deluxetable}. Make sure there is at least one \tablenotemark
%% in the table for each \tablenotetext.

\tablenotetext{*}{The NIFS On-Instrument WaveFront Sensor (OIWFS) is used to correct spatial flexure in the observations.}
\end{deluxetable}

\clearpage

\begin{deluxetable}{lccccccccc}
\tabletypesize{\scriptsize}
\tablecaption{Stellar Parameters and Mass Accretion Rates \label{tbl-2}}
\tablewidth{0pt}
\tablehead{
\colhead{Star} & \colhead{Mass} & \colhead{Luminosity} & \colhead{T$_{eff}$} & 
\colhead {A$_v$} & \colhead{Integrated Br$\gamma$ Flux} & \colhead {$\dot{M}_{acc}$} & 
\colhead{$\dot{M}_{out}$} & \colhead{Reference} \\

\colhead{}   &  \colhead{M$_{\odot}$} &  \colhead{L$_{\odot}$} &  \colhead{K} &  \colhead{mag} &  \colhead{W/m$^2$} & \colhead{M$_{\odot}$ yr$^{-1}$}  & \colhead{M$_{\odot}$ yr$^{-1}$}  & \colhead{}\\
\colhead{(1)}   &  \colhead{(2)} &  \colhead{(3)} &  \colhead{(4)} &  \colhead{(5)} &  \colhead{(6)} & \colhead{(7)}  & \colhead{(8)}  & \colhead{(9)}\\}

\startdata
CW Tau & 1.1 & 0.7 & 4700 &  2.2 &  1.5$\times10^{-16}$ & 1.3$\times10^{-08}$  & \nodata &  \\
%CW Tau & 1.1M$_{\odot}$ & 0.7L$_{\odot}$ & 4700K &  2.2 &  1.5e$^{-16}$ W/m$^2$ & 1.3e$^{-08}$M$_{\odot}$  & -- & \\
DG Tau & 2.2 & 7.7 & 4775 & 3.3  & 4.5$\times10^{-16}$  & 9.6$\times10^{-08}$  & $>$1.2$\times10^{-08}$ & \\
%DG Tau & 2.2 & 7.7 & 4775 & 3.3  & 4.5e$^{-16}$  & 9.6e$^{-08}$  & $>$1.2e$^{-08}$M$_{\odot}$ & \\
Haro 6-10 S & 0.7 & 1.8 & 4000 & 12.1 & 7.2$\times10^{-17}$ &  6.7$\times10^{-08}$ & $>$1.4$\times10^{-08}$ & Doppmann et al. \\ 
HL Tau & 1.2 & 3.0 & 4400 & 7.4  & 2.1$\times10^{-16}$ &  8.7$\times10^{-08}$ & \nodata & \\
RW Aur A & 2.8 & 12.9 & 5000 & 1.6  & 7.2$\times10^{-16}$ & 1.4$\times10^{-07}$ & \nodata & \\
RW Aur B & 1.2 & 3.0 & 4200 & 1.6  &  $<$5.1$\times10^{-18}$ & $<$4.1$\times10^{-10}$ & \nodata  & \\
T Tau N & 2.1 & 7.2 & 5250 & 1.5  &  8.4$\times10^{-16}$ & 1.5$\times10^{-07}$ & \nodata  & \\
XZ Tau A & 0.45 & 0.4 & 3400 & 1.4 & $<$7.7$\times10^{-18}$ & $<$1.9$\times10^{-09}$ & \nodata & \\

XZ Tau B & & & &  &  &  & & \\
Haro 6-10 N & 0.6 & 0.8 & 4100 & 29.0  & 1.7$\times10^{-17}$ & 4.6$\times10^{-08}$ & \nodata & Doppmann et al. \\ 

\enddata
%% Text for table notes should follow after the \enddata but before
%% the \end{deluxetable}. Make sure there is at least one \tablenotemark
%% in the table for each \tablenotetext.

\end{deluxetable}

\clearpage

\begin{figure}
%\plotfiddle{epsfile}{vsize}{rotation}{hscale}{vscale}{htrans}{vtrans}
%example:
\plotfiddle{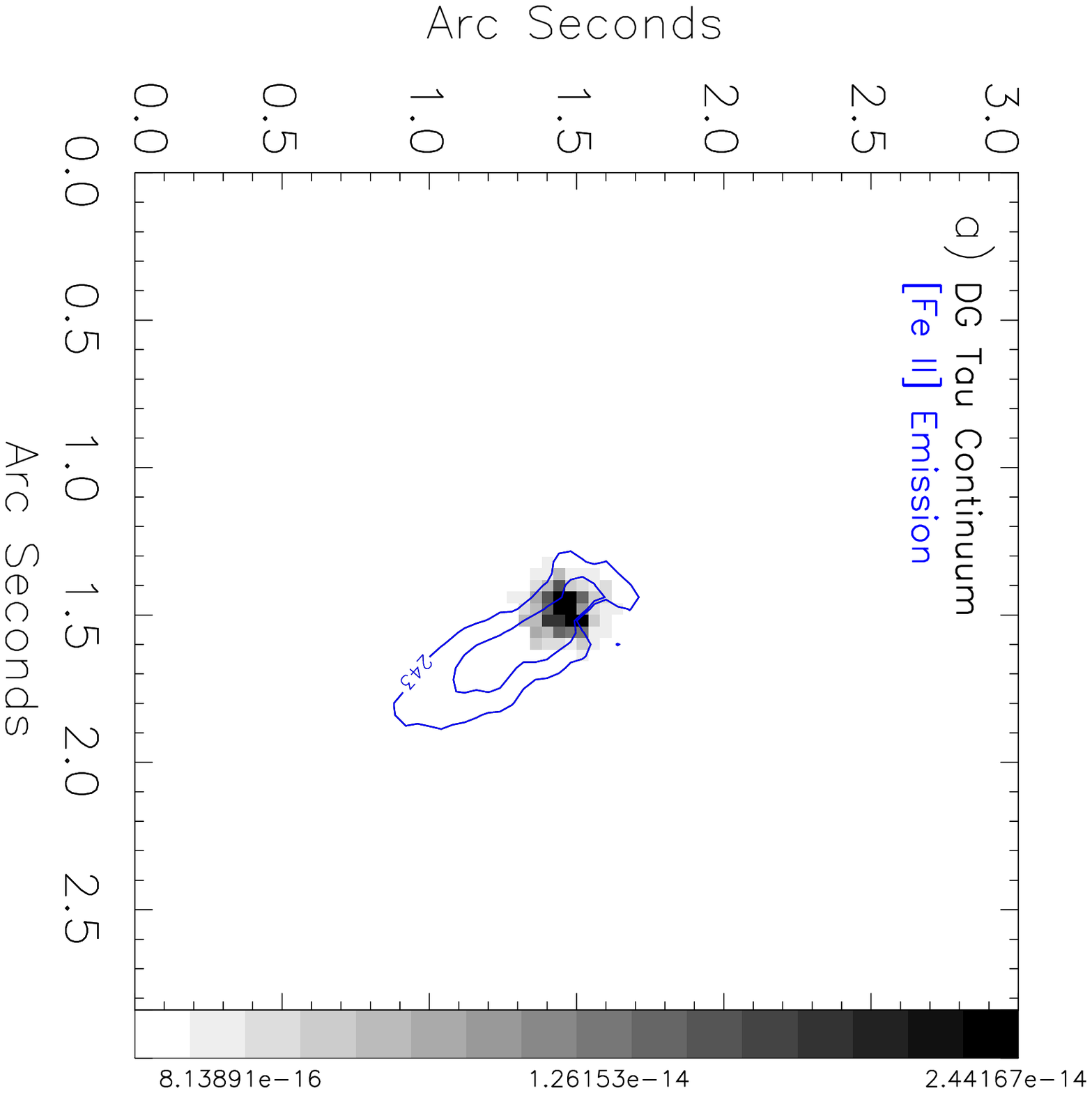}{350pt}{90pt}{50pt}{50pt}{70pt}{100pt}
\plotfiddle{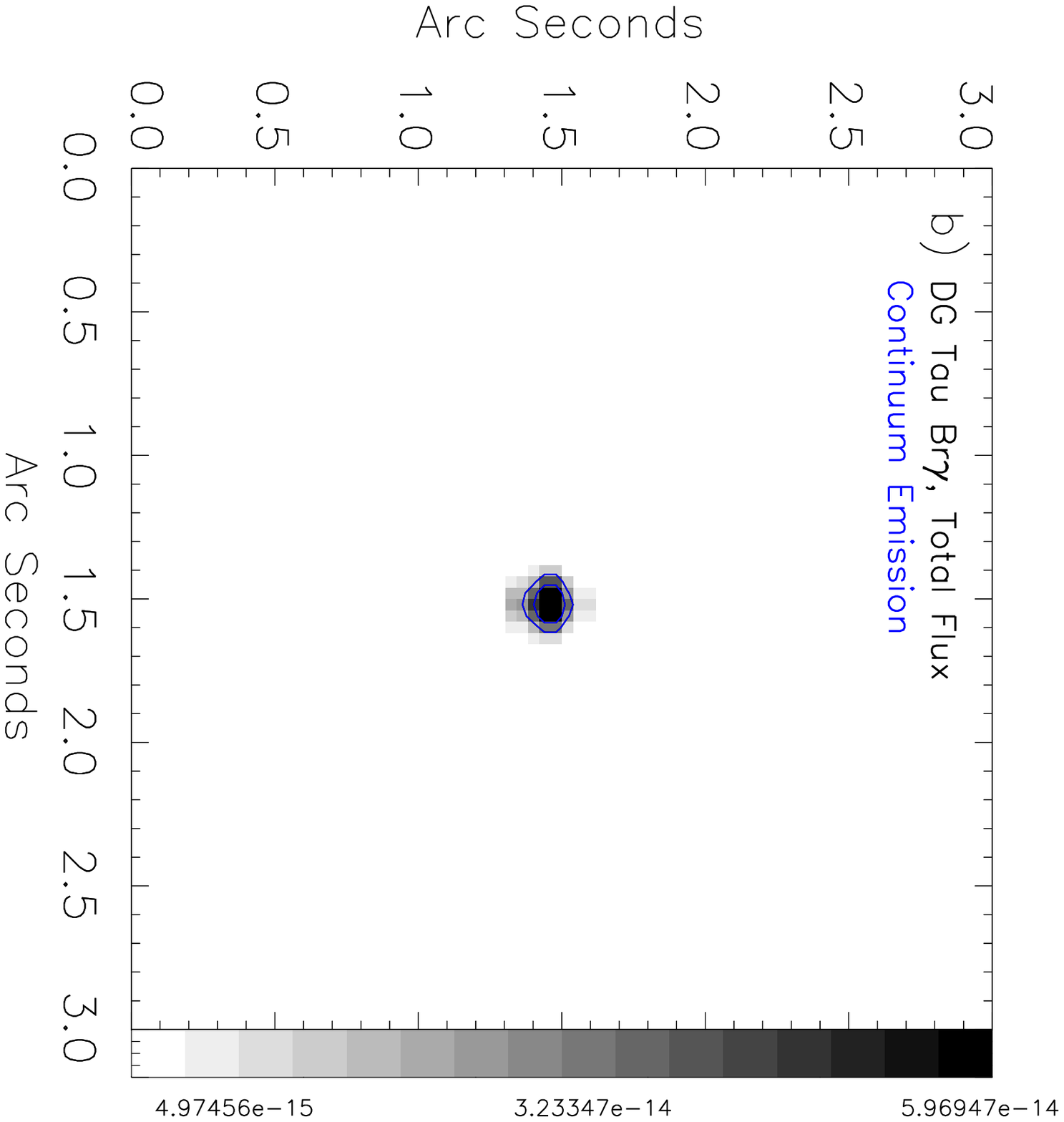}{350pt}{90pt}{50pt}{50pt}{330pt}{466pt}
\plotfiddle{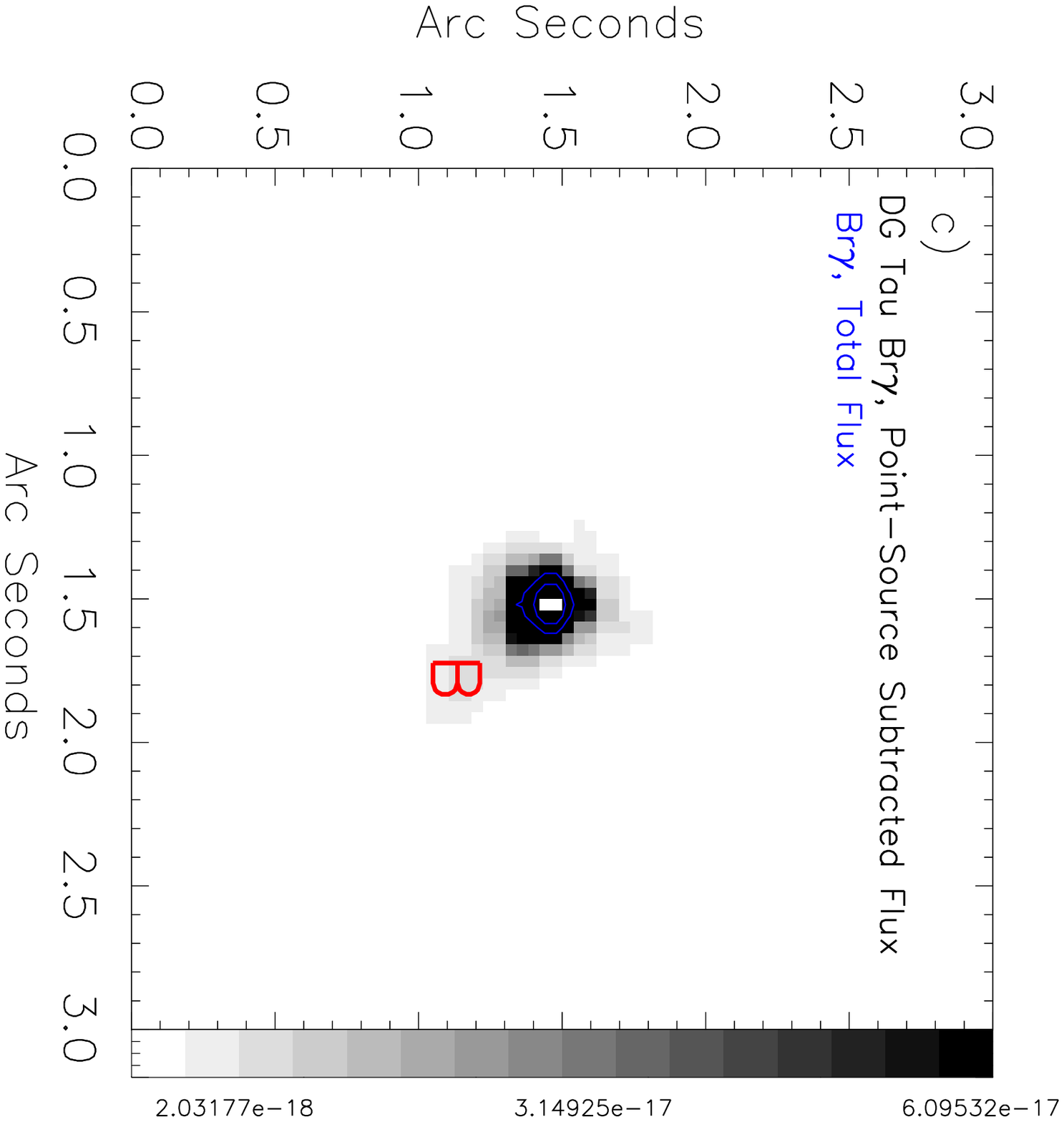}{350pt}{90pt}{50pt}{50pt}{70pt}{600pt}
\plotfiddle{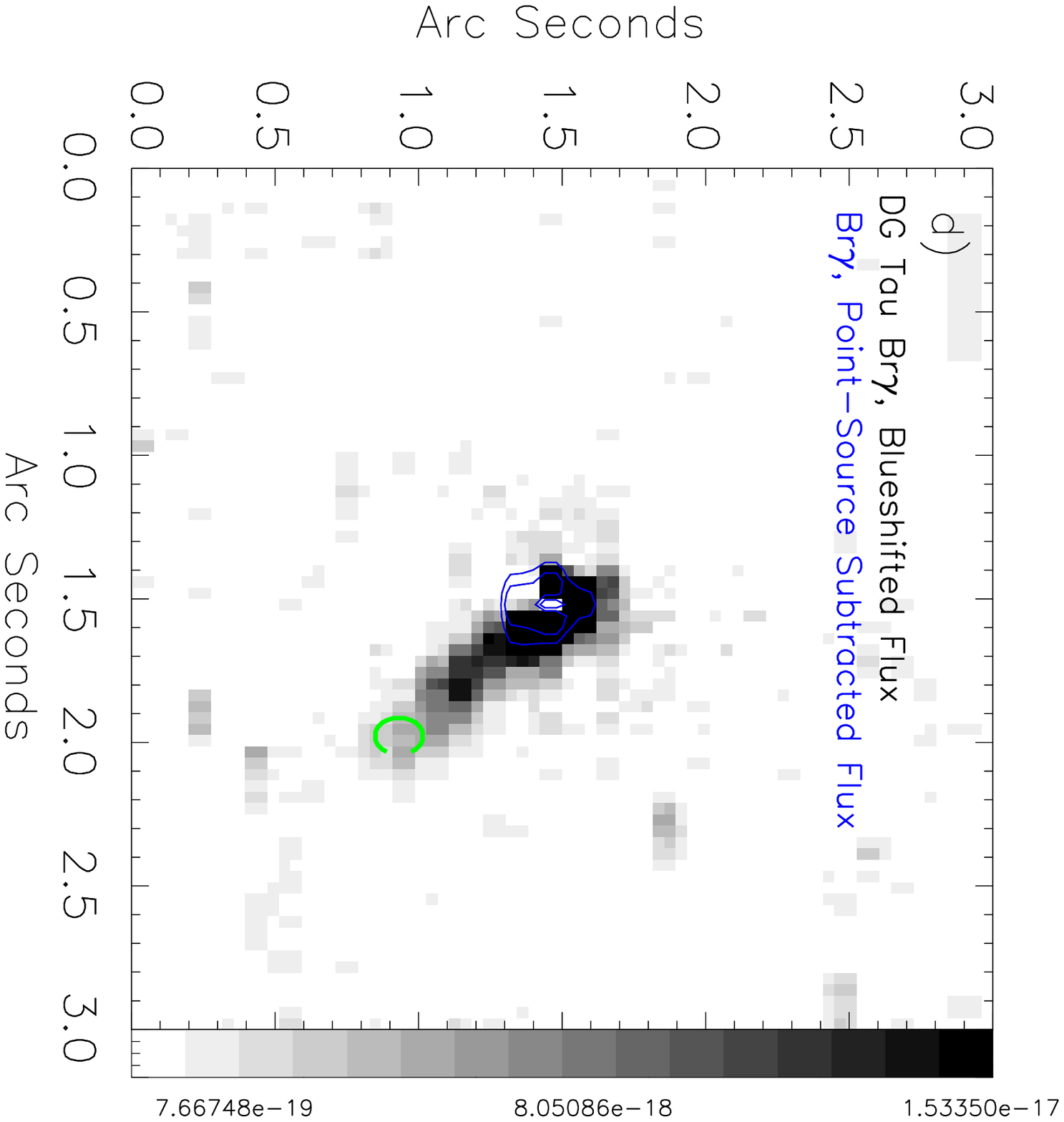}{350pt}{90pt}{50pt}{50pt}{330pt}{966pt}
\vskip -13.5in
\caption{(a) The 2.16 $\mu$m continuum flux level from DG Tau with contours of [Fe II] emission overplotted, designating the outflow location, (b) the continuum subtracted point-source Br$\gamma$ flux and (c) the continuum and point-source subtracted map of spatially extended Br$\gamma$ from DG Tau.  Also included here is a continuum and point-source subtracted map of spatially extended Br$\gamma$, integrated over only three blue-shifted velocity channels (d).  The (d) panels show that much of the spatially extended Br$\gamma$ emission is blue-shifted. The locations designated as "B" and "C" had 1-D spectral traces extracted, these are presented in Figure 5.}
\end{figure}

\clearpage

\begin{figure}
%\plotfiddle{epsfile}{vsize}{rotation}{hscale}{vscale}{htrans}{vtrans}
%example:
\vskip -1.5in
\plotfiddle{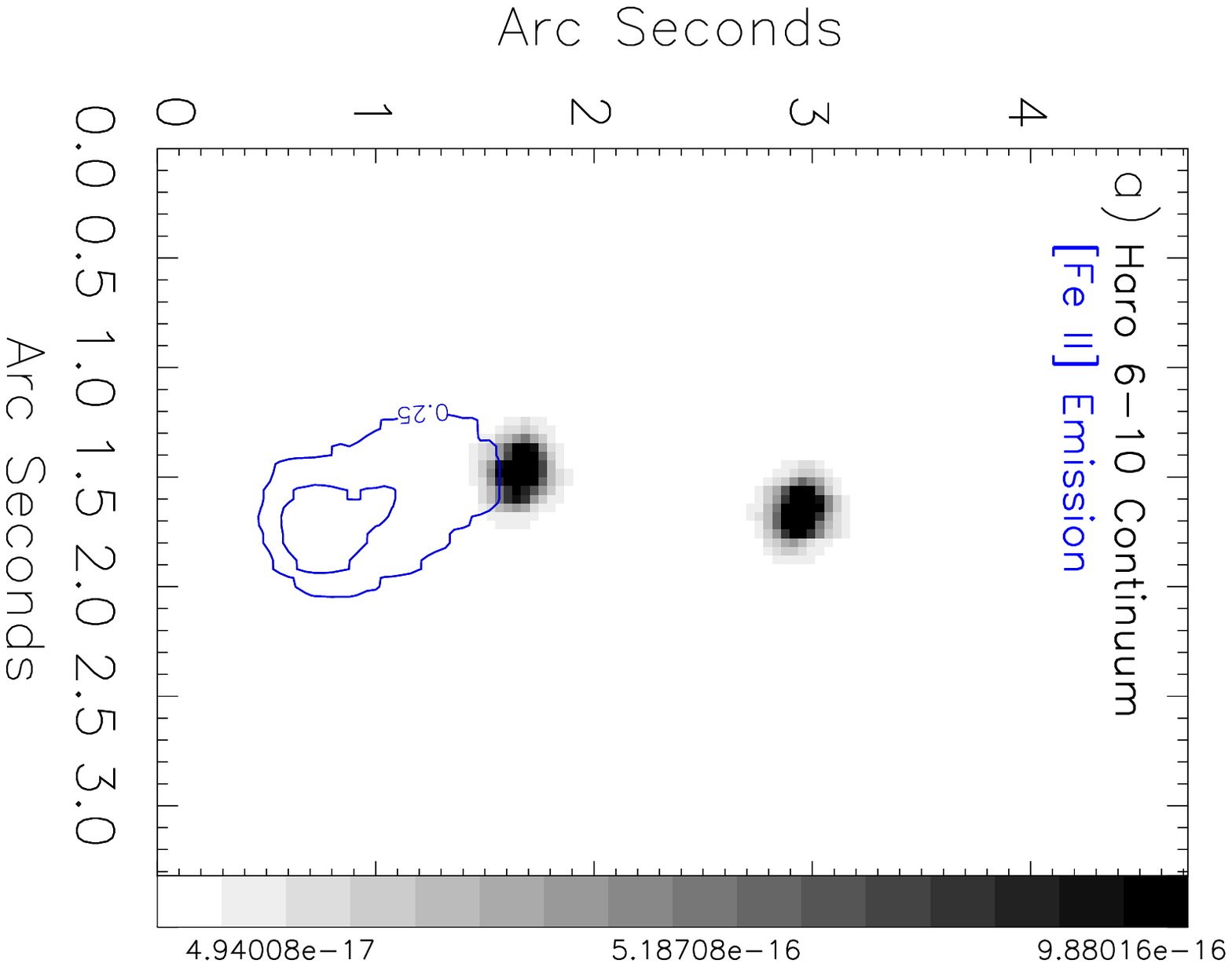}{350pt}{90pt}{70pt}{70pt}{150pt}{30pt}
\plotfiddle{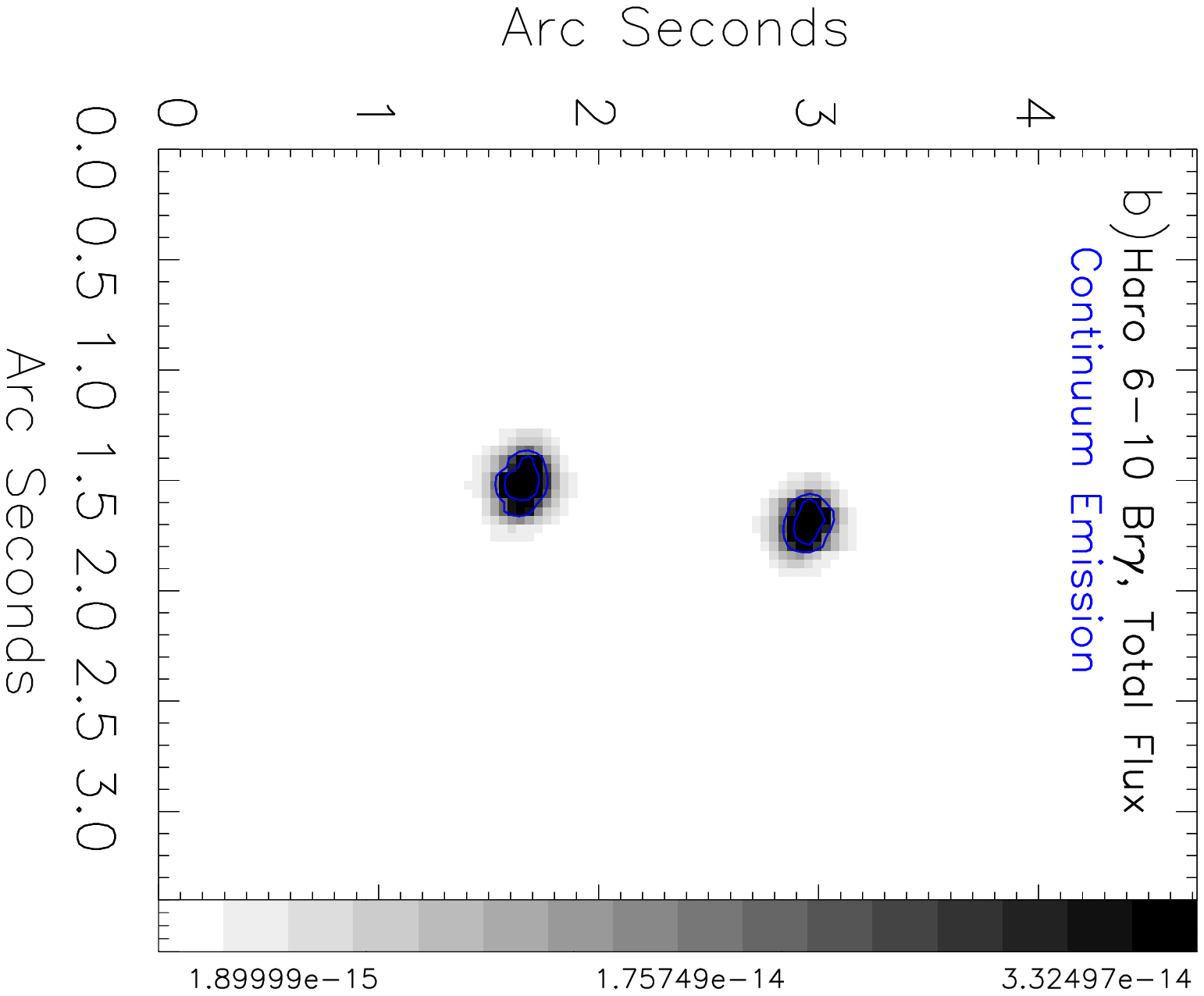}{350pt}{90pt}{70pt}{70pt}{415pt}{395pt}
\plotfiddle{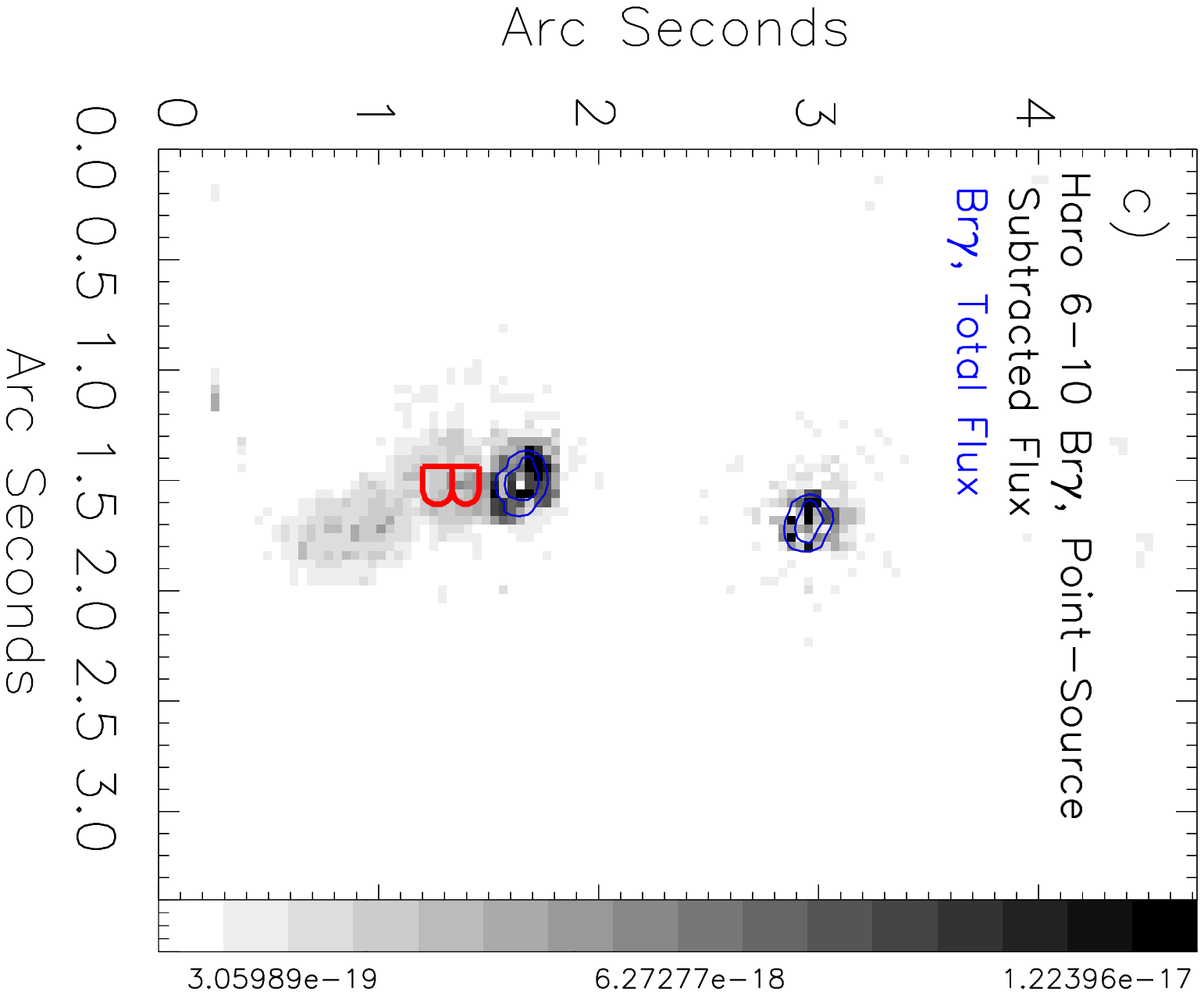}{350pt}{90pt}{70pt}{70pt}{150pt}{450pt}
\plotfiddle{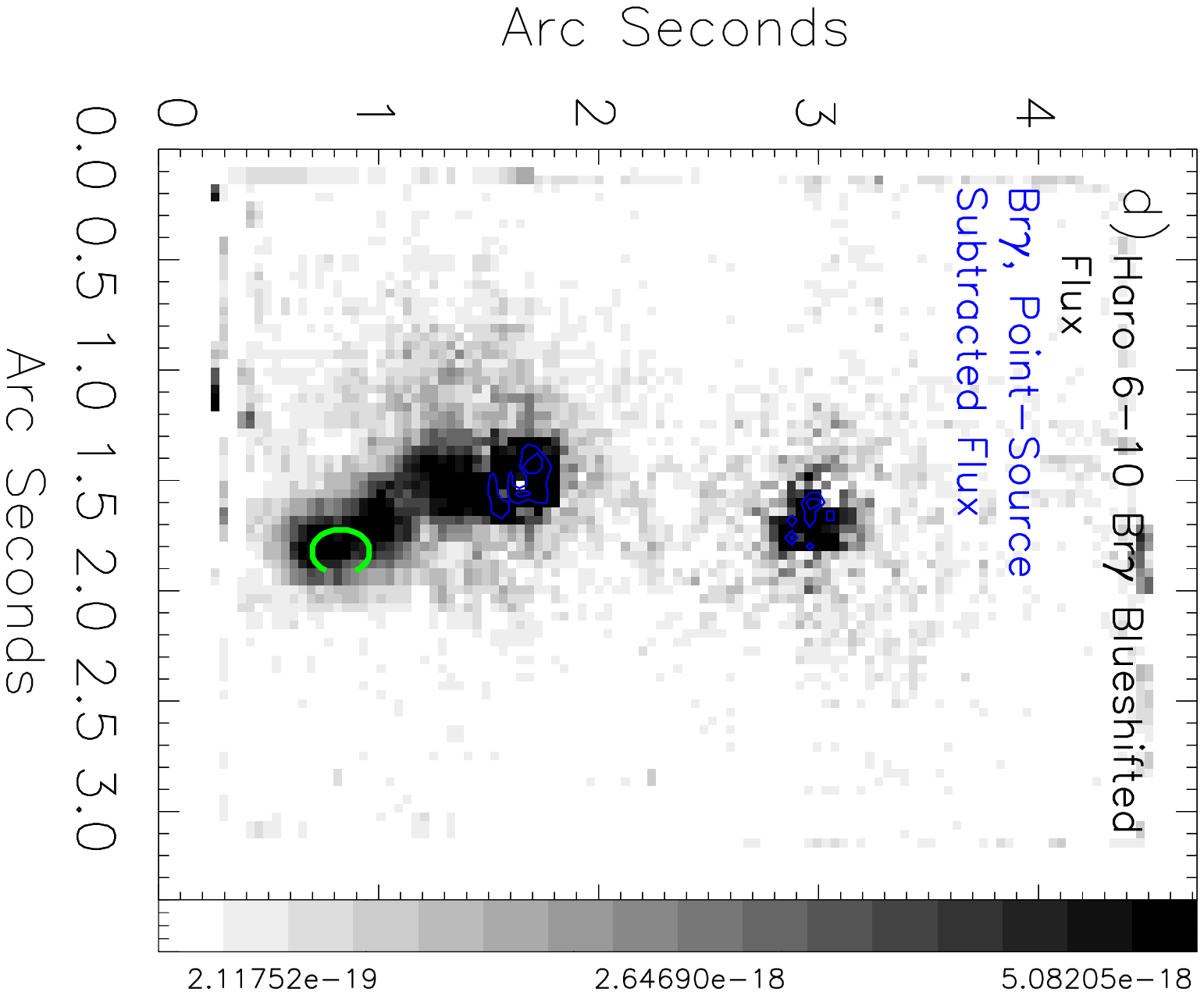}{350pt}{90pt}{70pt}{70pt}{415pt}{815pt}
\vskip -11.5in
\caption{(a) The 2.16 $\mu$m continuum flux level from Haro 6-10 with contours of [Fe II] emission overplotted, designating the outflow location, (b) the continuum subtracted point-source Br$\gamma$ flux and (c) the continuum and point-source subtracted map of spatially extended Br$\gamma$ from Haro 6-10 S.  Also included here is a continuum and point-source subtracted map of spatially extended Br$\gamma$, integrated over only three blue-shifted velocity channels (d).  The (d) panels show that much of the spatially extended Br$\gamma$ emission is blue-shifted.  The locations designated as "B" and "C" had 1-D spectral traces extracted, these are presented in Figure 5.}
\end{figure}

\begin{figure}
%\plotfiddle{epsfile}{vsize}{rotation}{hscale}{vscale}{htrans}{vtrans}
%example:
\plotfiddle{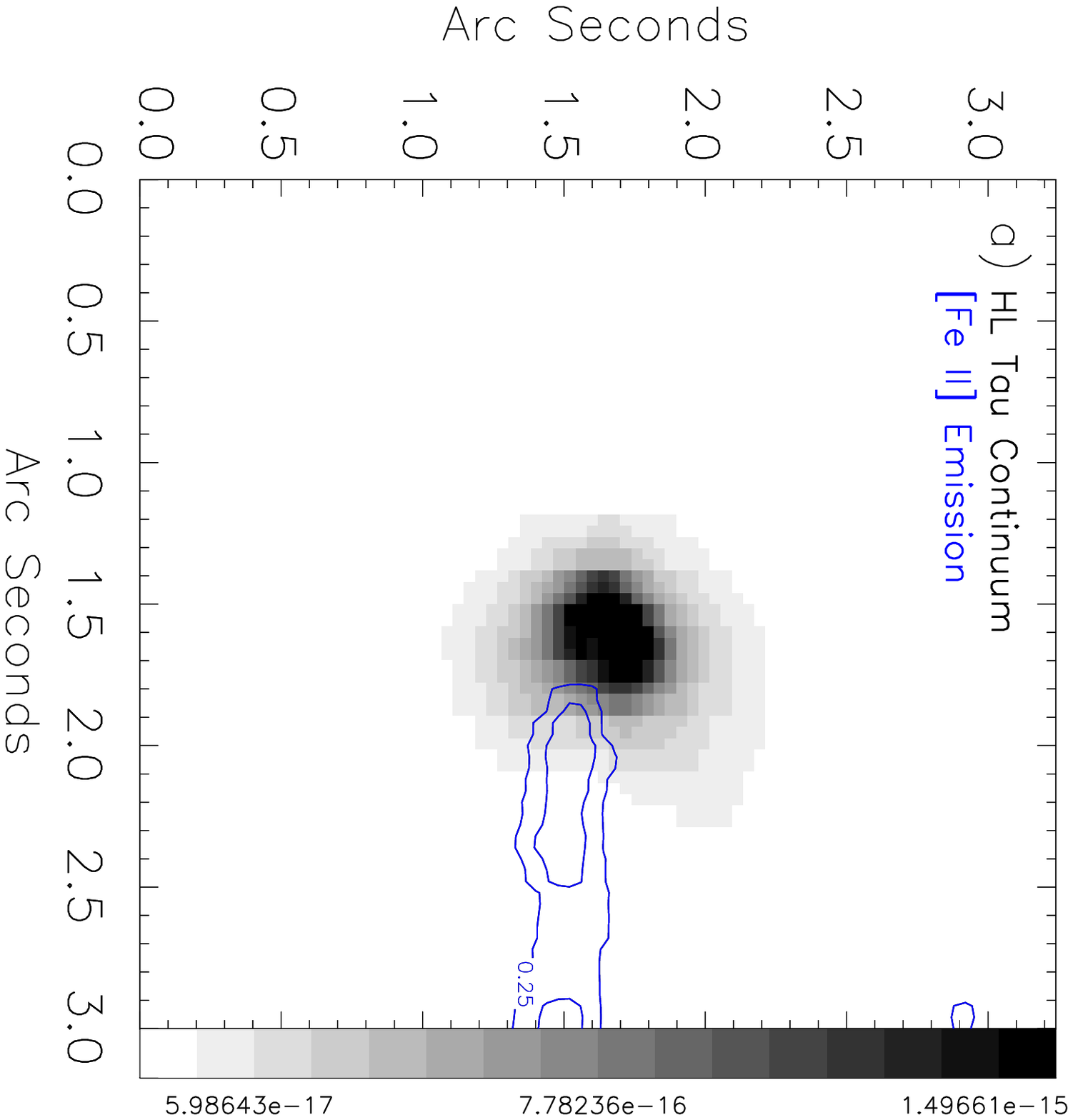}{350pt}{90pt}{50pt}{50pt}{70pt}{100pt}
\plotfiddle{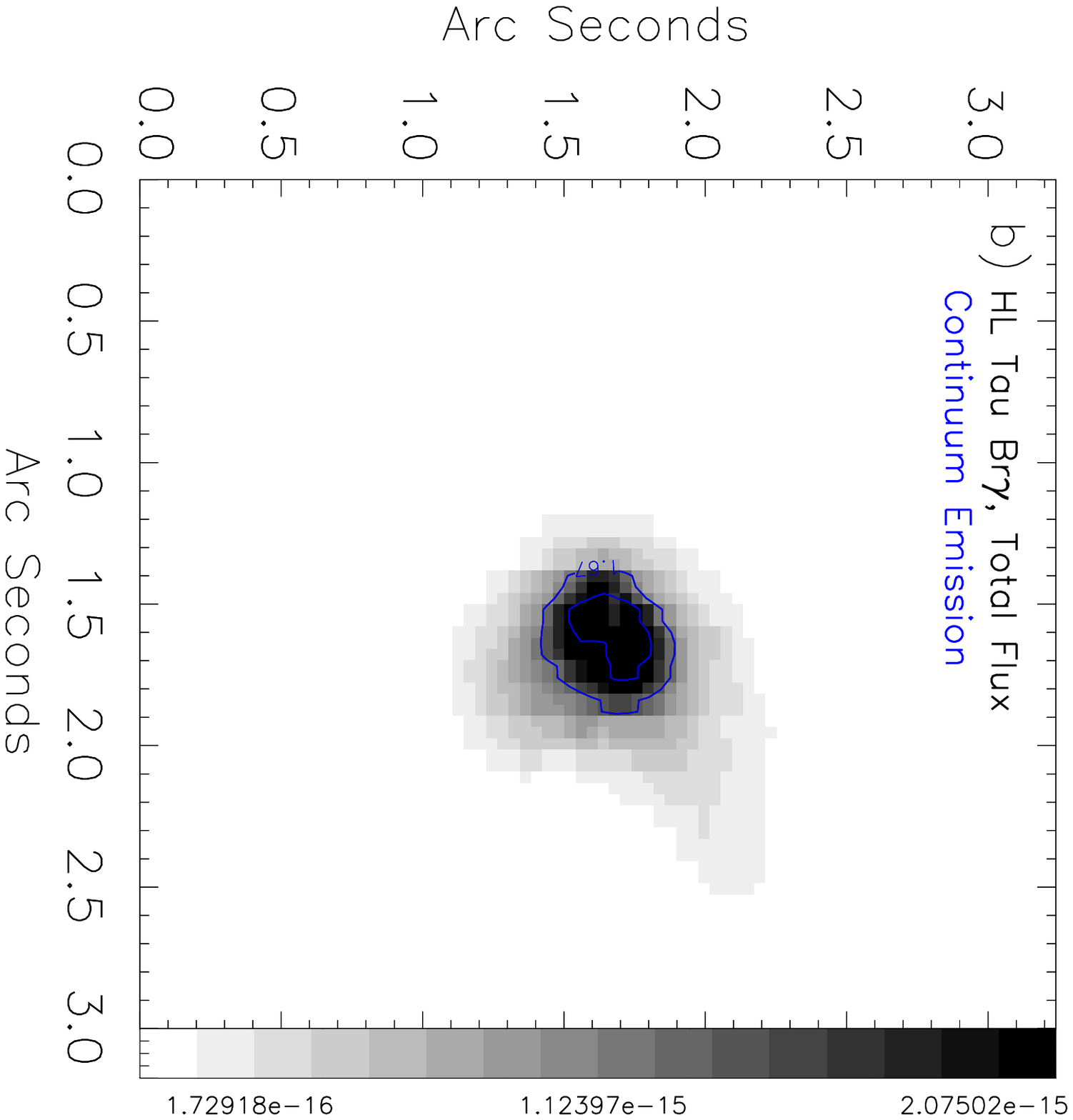}{350pt}{90pt}{50pt}{50pt}{330pt}{466pt}
\plotfiddle{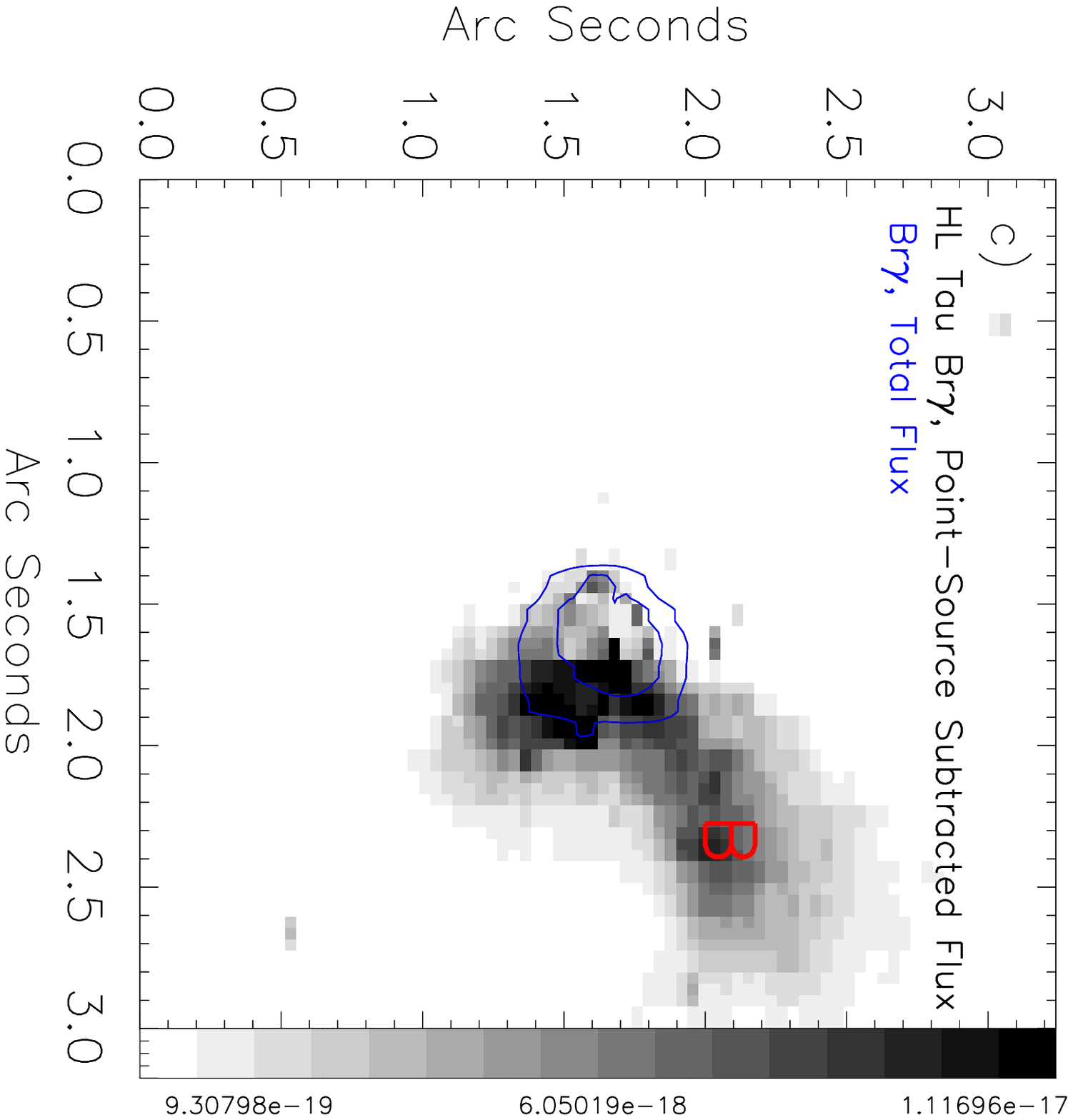}{350pt}{90pt}{50pt}{50pt}{70pt}{600pt}
\plotfiddle{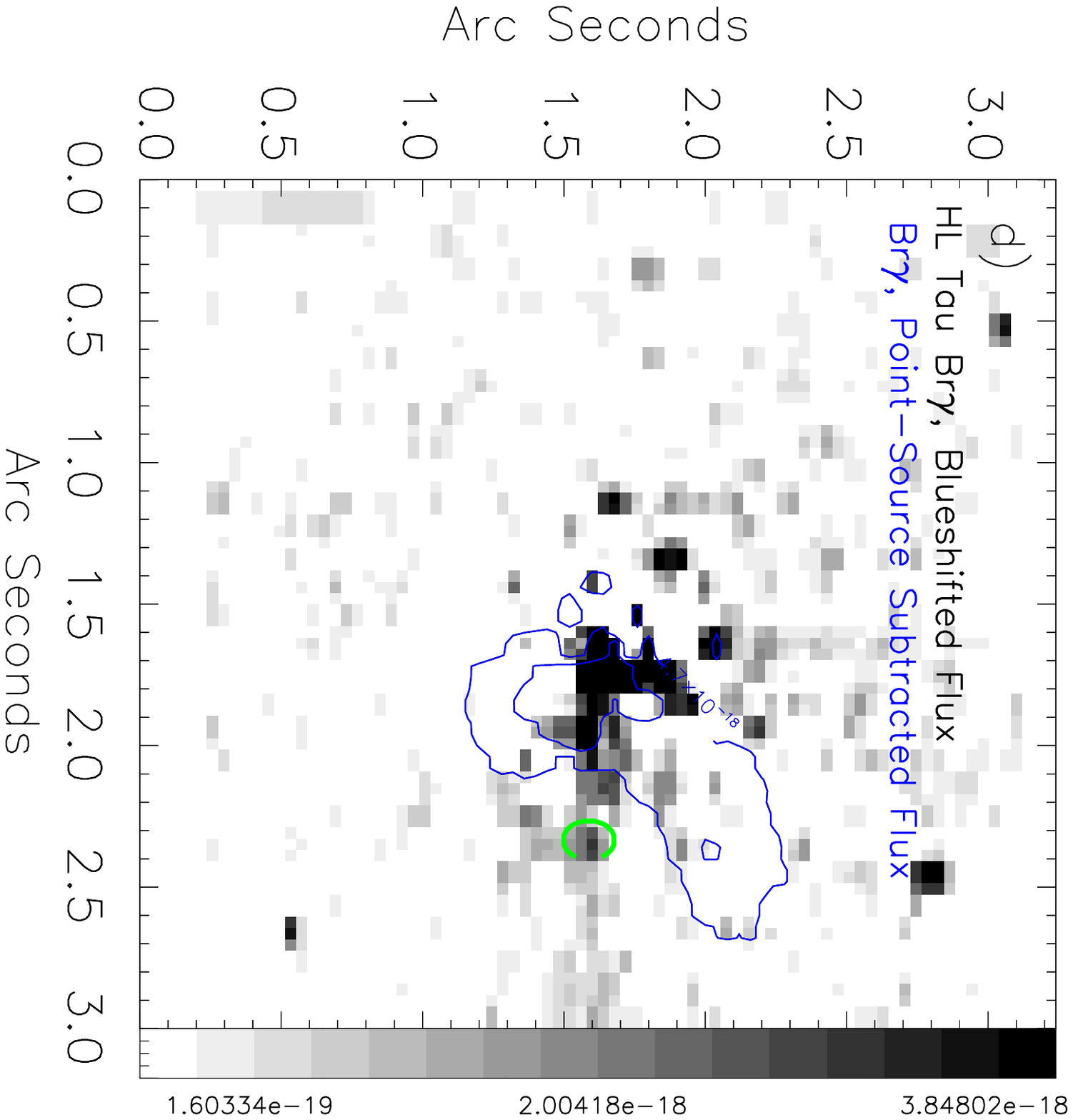}{350pt}{90pt}{50pt}{50pt}{330pt}{966pt}
\vskip -13.5in
\caption{(a) The 2.16 $\mu$m continuum flux level from HL Tau with the 0.$"$2 occulting disk in the beam  with contours of [Fe II] emission overplotted, designating the outflow location, (b) the continuum subtracted point-source Br$\gamma$ flux and (c) the continuum and (d) point-source subtracted map of spatially extended Br$\gamma$ from HL Tau.  Also included here is a continuum and point-source subtracted map of spatially extended Br$\gamma$, integrated over only two blue-shifted velocity channels (d).   The (d) panel shows weak blue-shifted Br$\gamma$ emission that arises from the same spatial location as the [Fe II] emission from the outflow.  The locations designated as "B" and "C" had 1-D spectral traces extracted, these are presented in Figure 5.}
\end{figure}

\begin{figure}
%\plotfiddle{epsfile}{vsize}{rotation}{hscale}{vscale}{htrans}{vtrans}
%example:
\plotfiddle{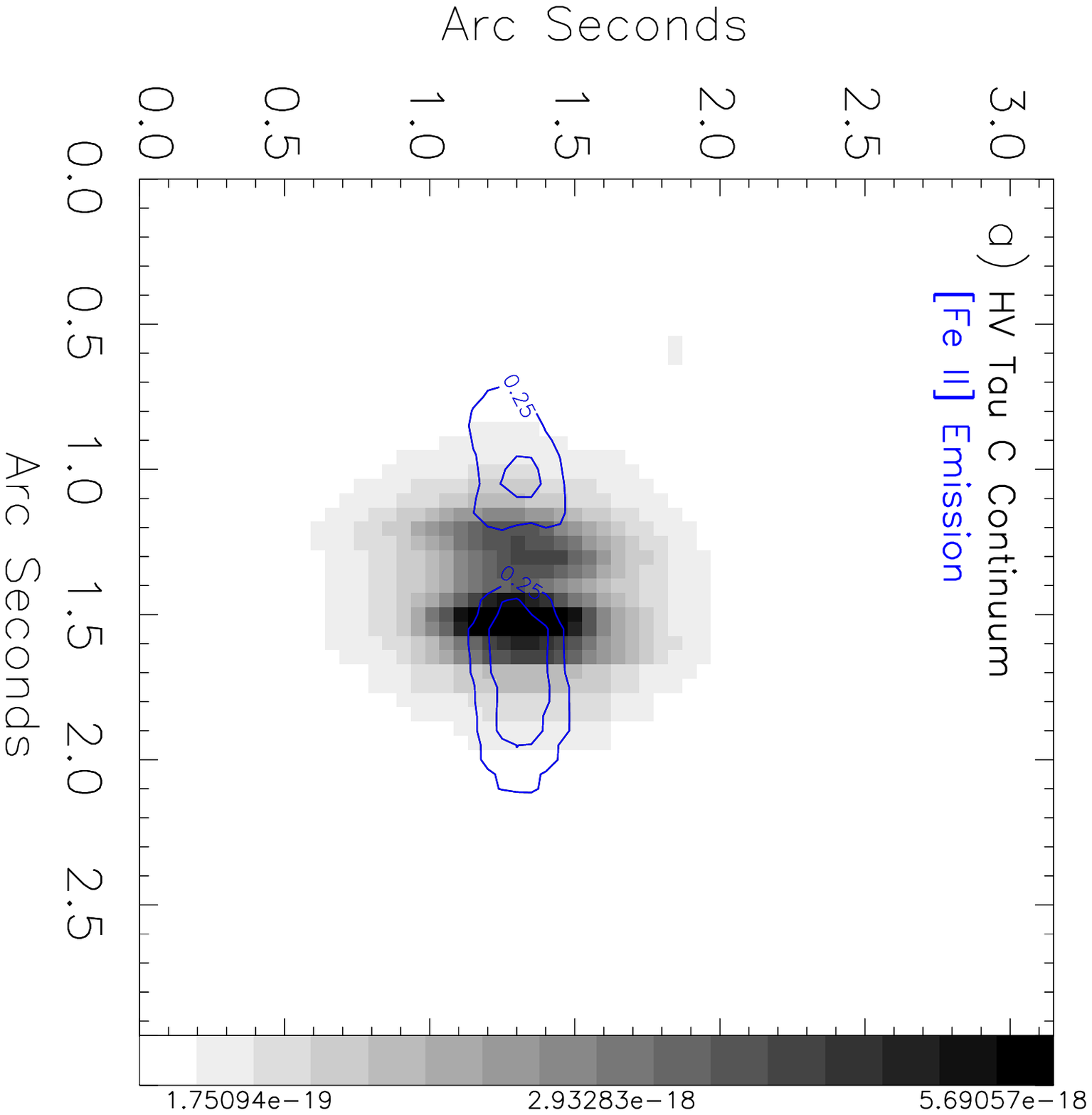}{350pt}{90pt}{50pt}{50pt}{70pt}{100pt}
\plotfiddle{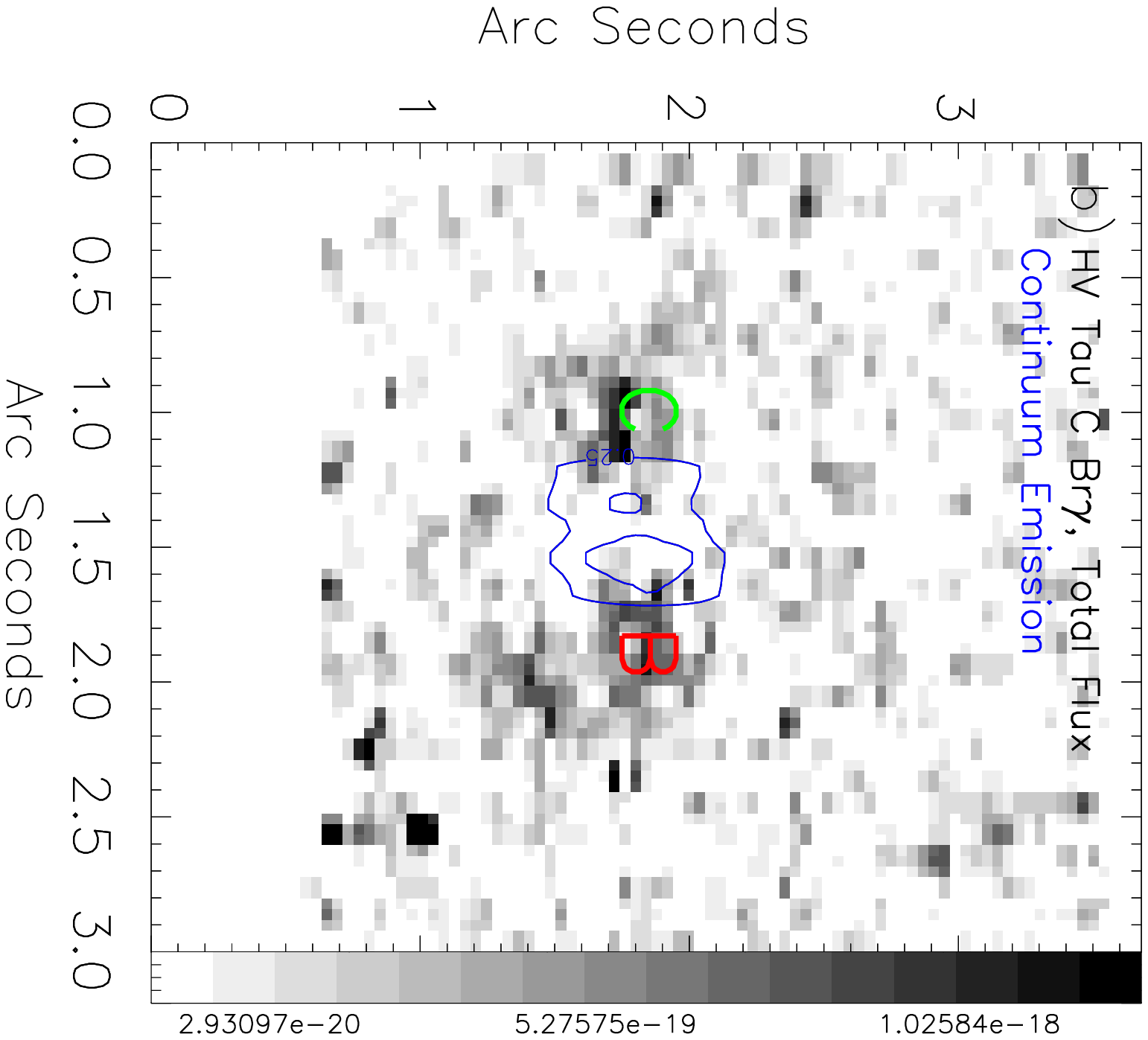}{350pt}{90pt}{58pt}{50pt}{350pt}{466pt}
\vskip -6.5in
\caption{(a) The 2.16 $\mu$m continuum flux level from HV Tau C with contours of [Fe II] emission overplotted, designating the outflow location, and (b) the continuum subtracted point-source Br$\gamma$ flux. The locations designated as "B" and "C" had 1-D spectral traces extracted, these are presented in Figure 5.}
\end{figure}

\begin{figure}
%\plotfiddle{epsfile}{vsize}{rotation}{hscale}{vscale}{htrans}{vtrans}
%example:
\plotfiddle{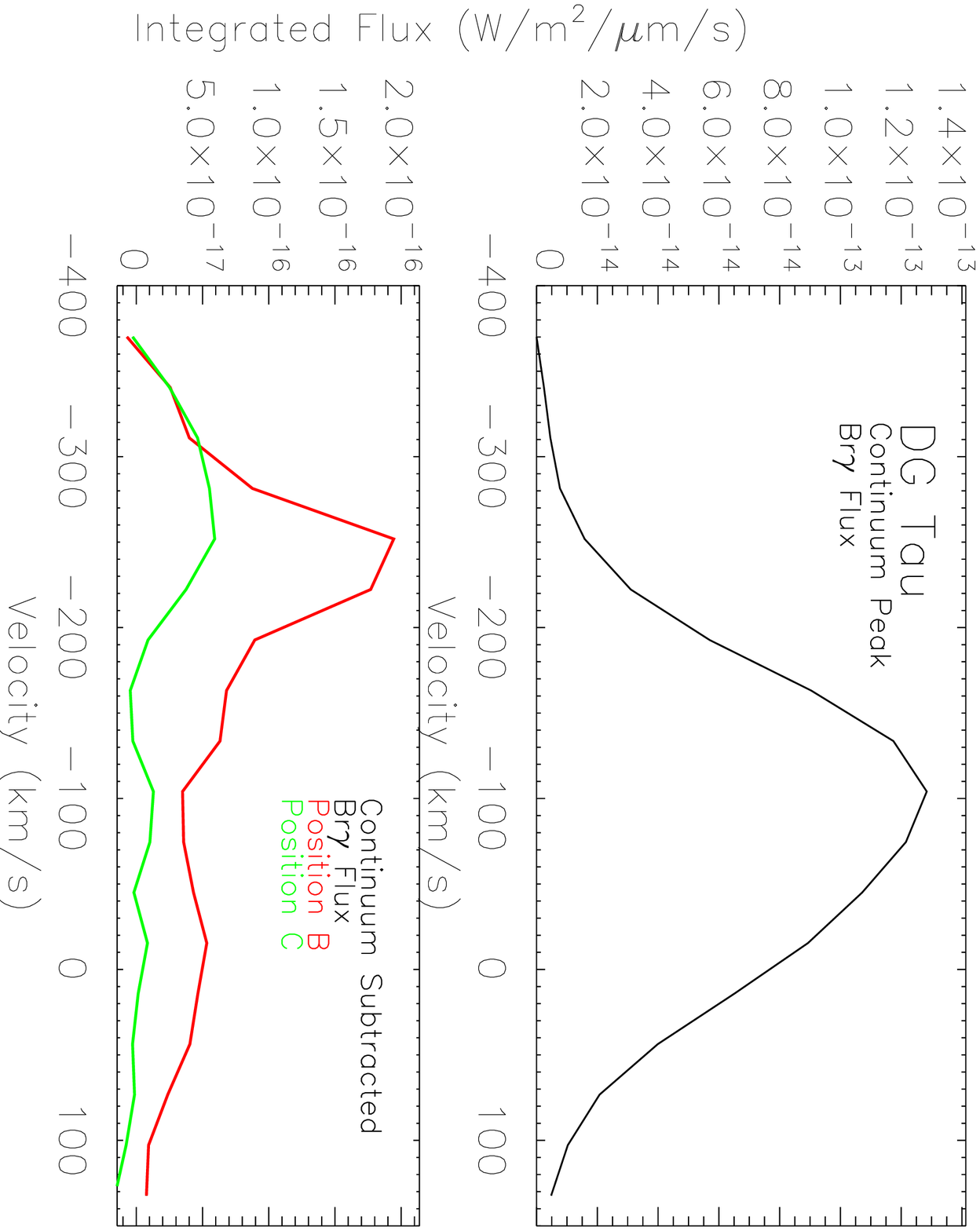}{350pt}{90pt}{40pt}{40pt}{10pt}{100pt}
\plotfiddle{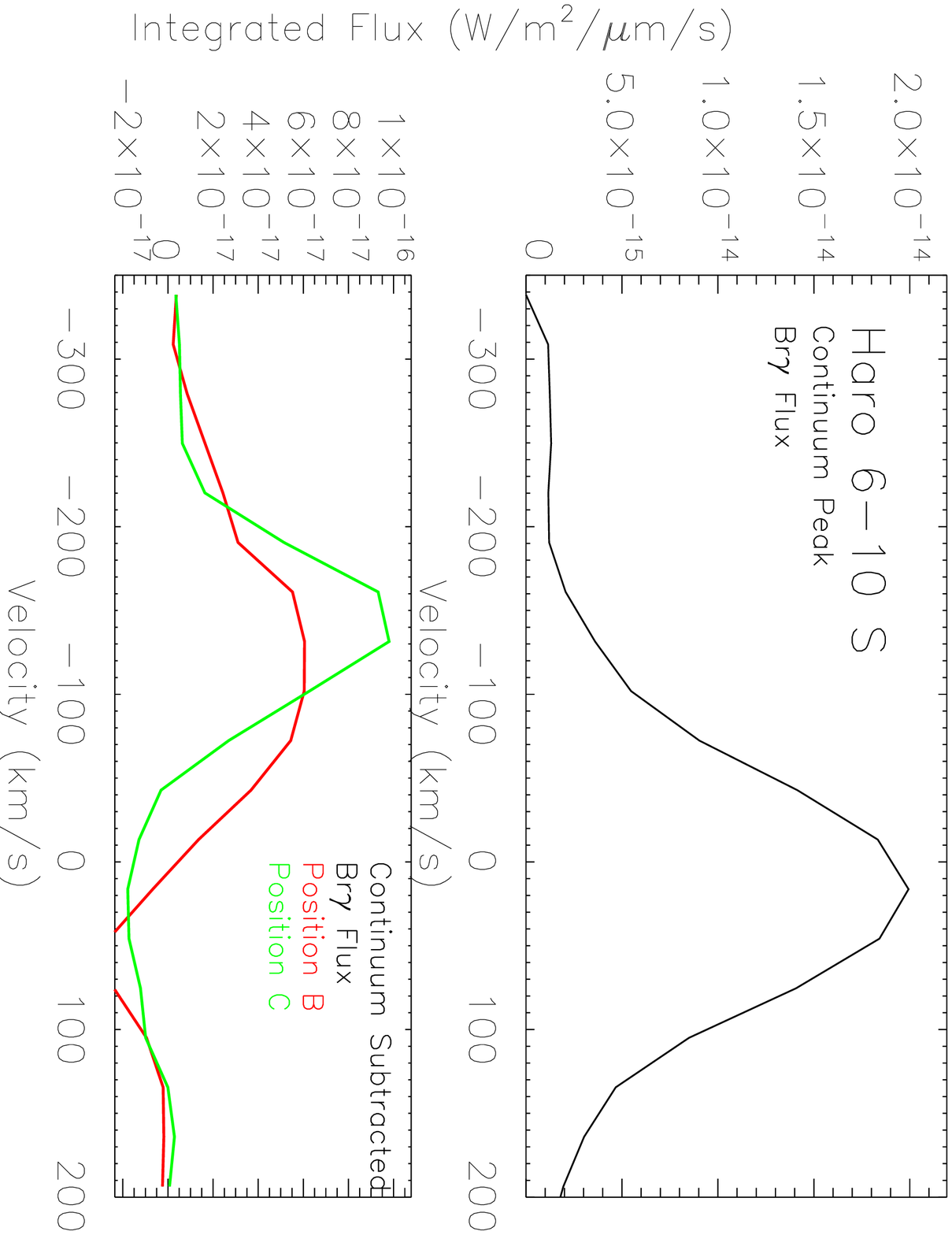}{350pt}{90pt}{40pt}{40pt}{290pt}{466pt}
\plotfiddle{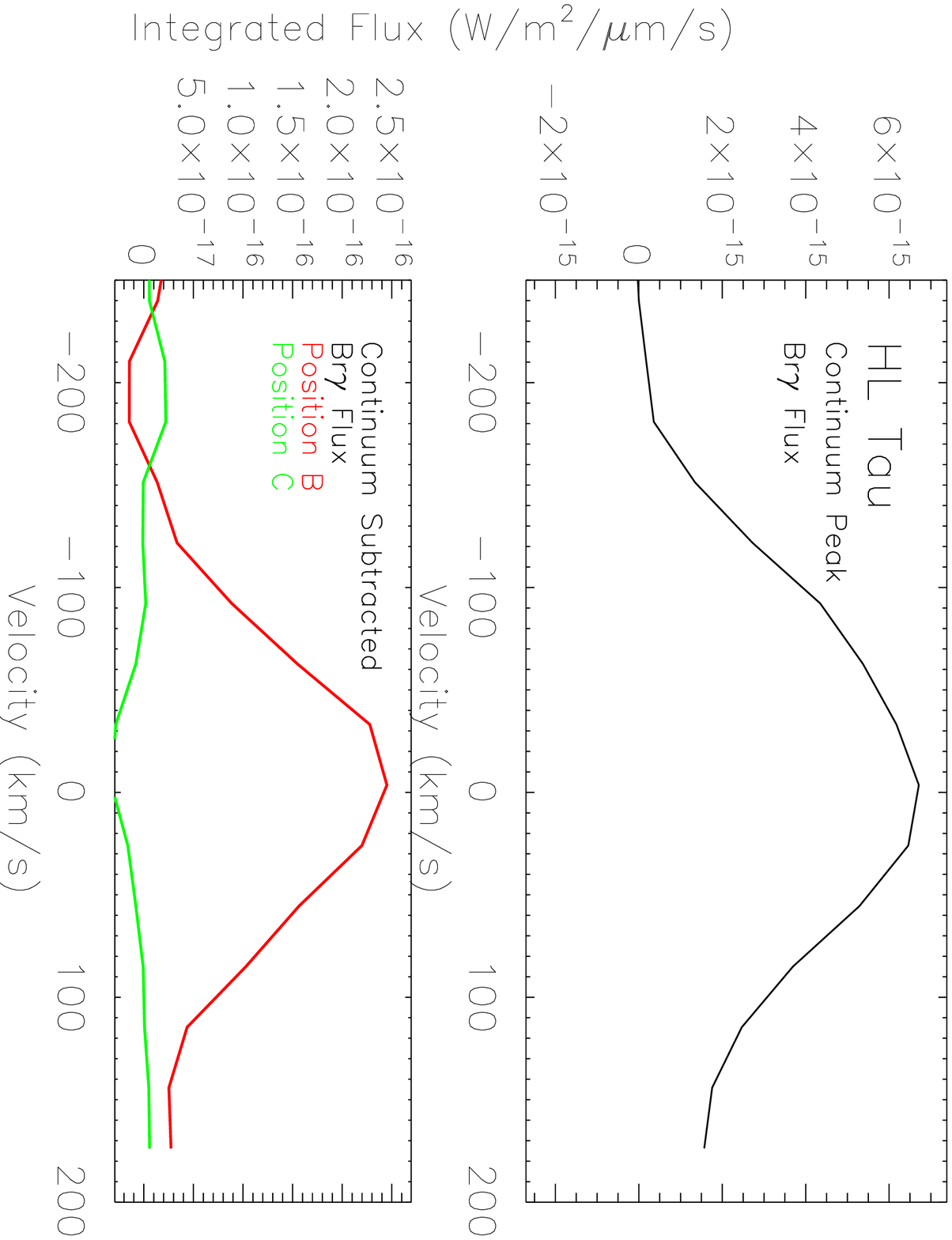}{350pt}{90pt}{40pt}{40pt}{10pt}{611pt}
\plotfiddle{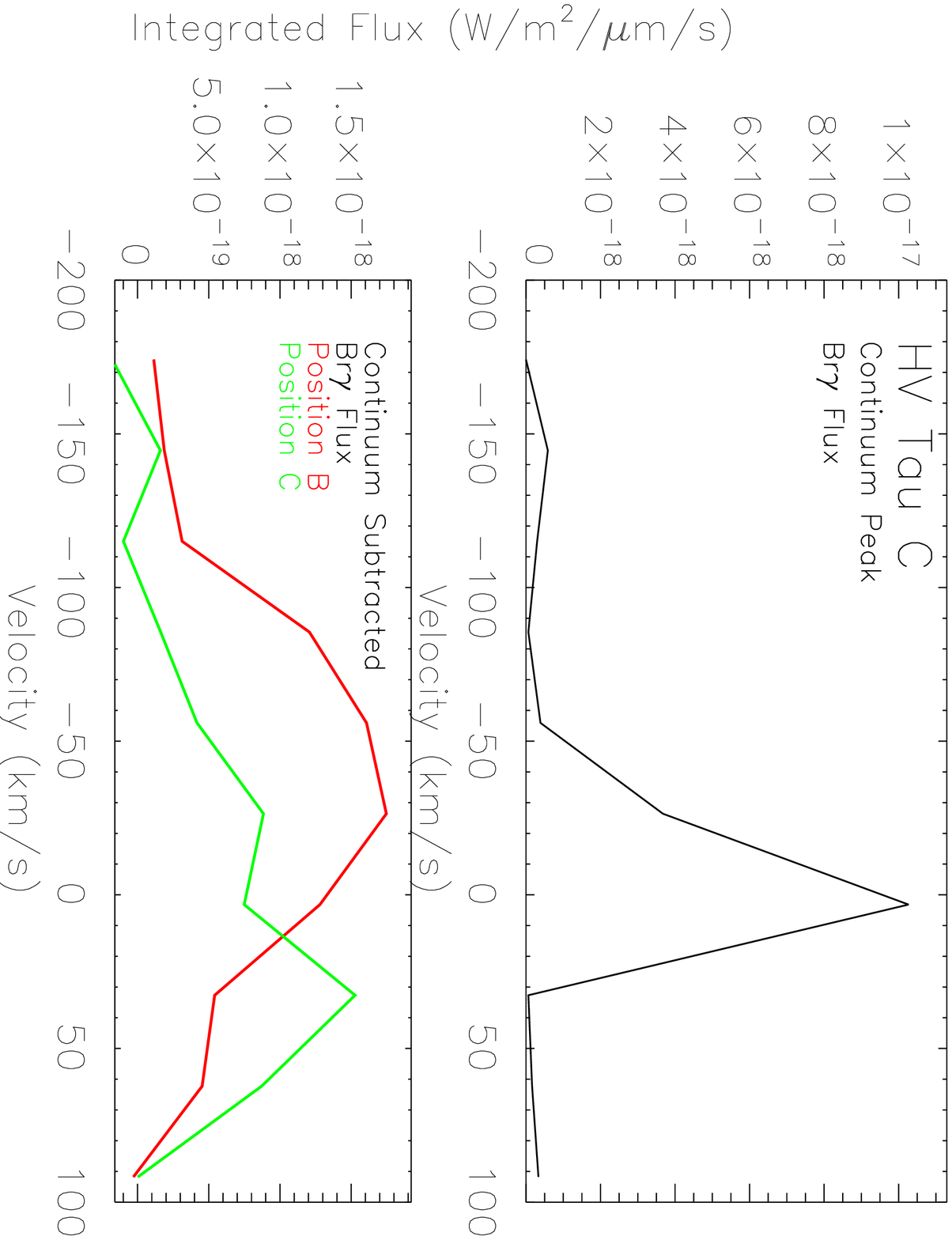}{350pt}{90pt}{40pt}{40pt}{290pt}{978pt}
\vskip -13.5in
\caption{Br$\gamma$ flux versus velocity plots at the continuum peak locations and "Position B" and "Position C" (from plots 1-4) for DG Tau, Haro 6-10 S, HL Tau and HV Tau C.  For DG Tau, HL Tau, and HV Tau C, the more spatially extended Br$\gamma$ emission measured at the "Position C" location is fainter and more blue-shifted than the less extended flux found at "Position B".  For Haro 6-10 S, the "Position C" flux measures a bright knot in the Herbig-Haro flow and is stronger and more blue-shifted than the "Position B" emission.}
\end{figure}

\begin{figure}
%\plotfiddle{epsfile}{vsize}{rotation}{hscale}{vscale}{htrans}{vtrans}
%example:
\plotfiddle{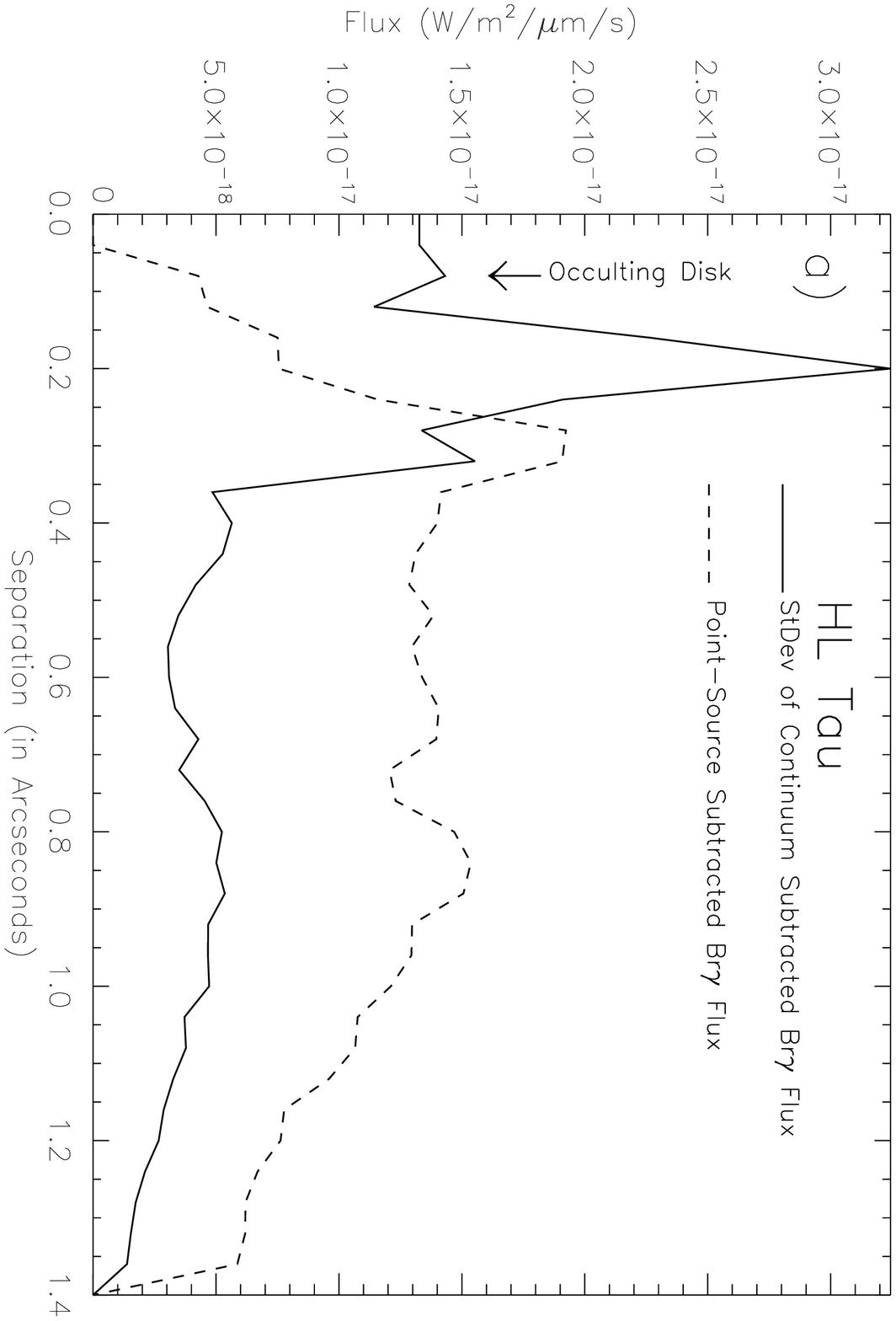}{350pt}{90pt}{40pt}{40pt}{20pt}{100pt}
\plotfiddle{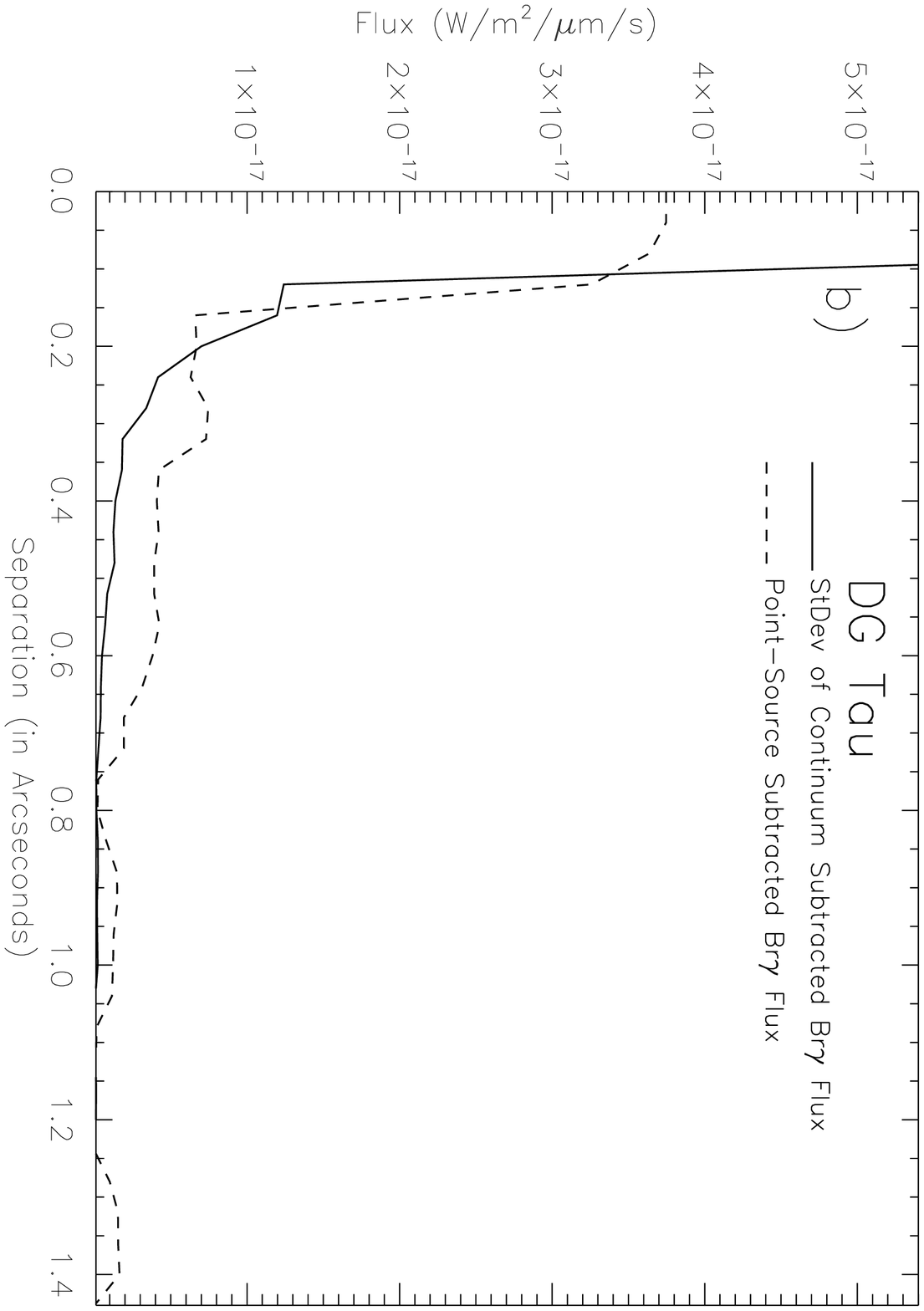}{350pt}{90pt}{40pt}{40pt}{300pt}{466pt}
\plotfiddle{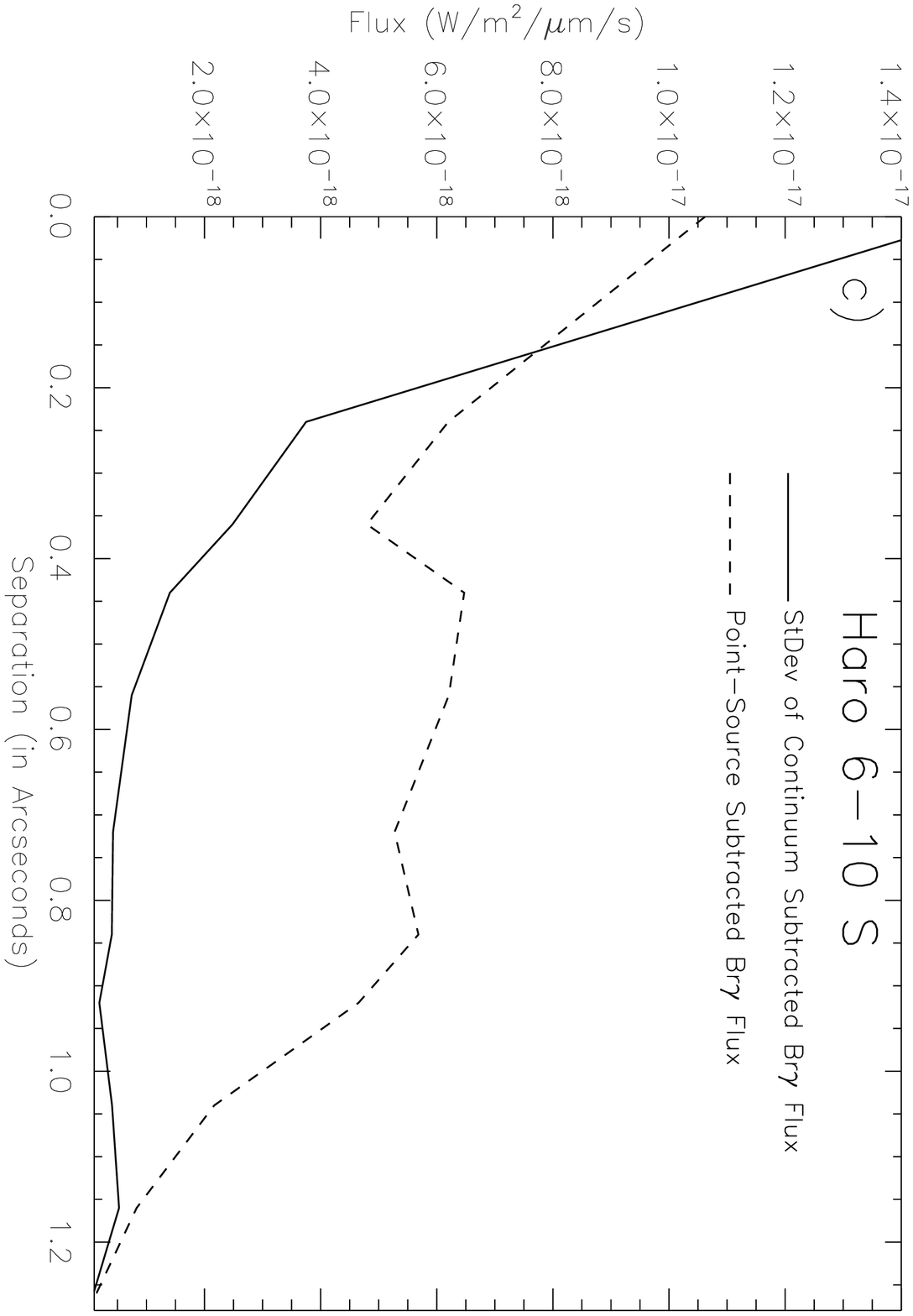}{350pt}{90pt}{40pt}{40pt}{20pt}{611pt}
\plotfiddle{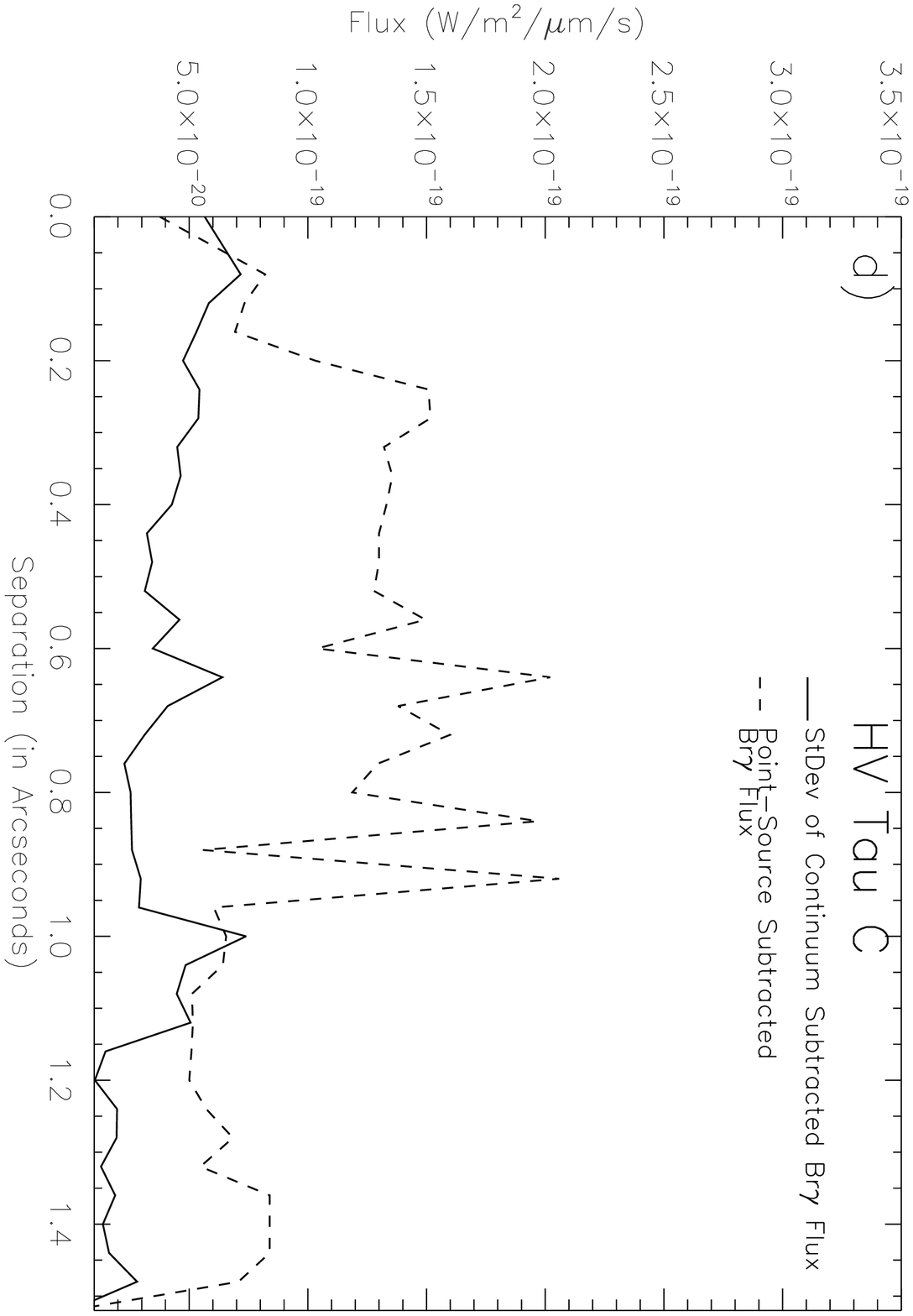}{350pt}{90pt}{40pt}{40pt}{300pt}{978pt}
\vskip -13.5in
\caption{The standard deviation of the continuum subtracted Br$\gamma$ flux plotted versus distance from the central continuum peak for HL Tau (a), DG Tau (b), Haro 6-10 S (c) and HV Tau C (d). The overplotted dashed line represents the level of spatially extended continuum and point-source subtracted Br$\gamma$ flux.  The ratio of the spatially extended flux to the standard deviation of point-source subtracted emission (dashed line to solid line) represents a measure of the signal-to-noise of the spatially extended Br$\gamma$.  The peak of the S/N for extended Br$\gamma$ is $\sim$11 for HL Tau, $\sim$6 for DG Tau, $\sim$25 for Haro 6-10 S, and $\sim$11 for HV Tau C.}
\end{figure}

\begin{figure}
%\plotfiddle{epsfile}{vsize}{rotation}{hscale}{vscale}{htrans}{vtrans}
%example:
\plotfiddle{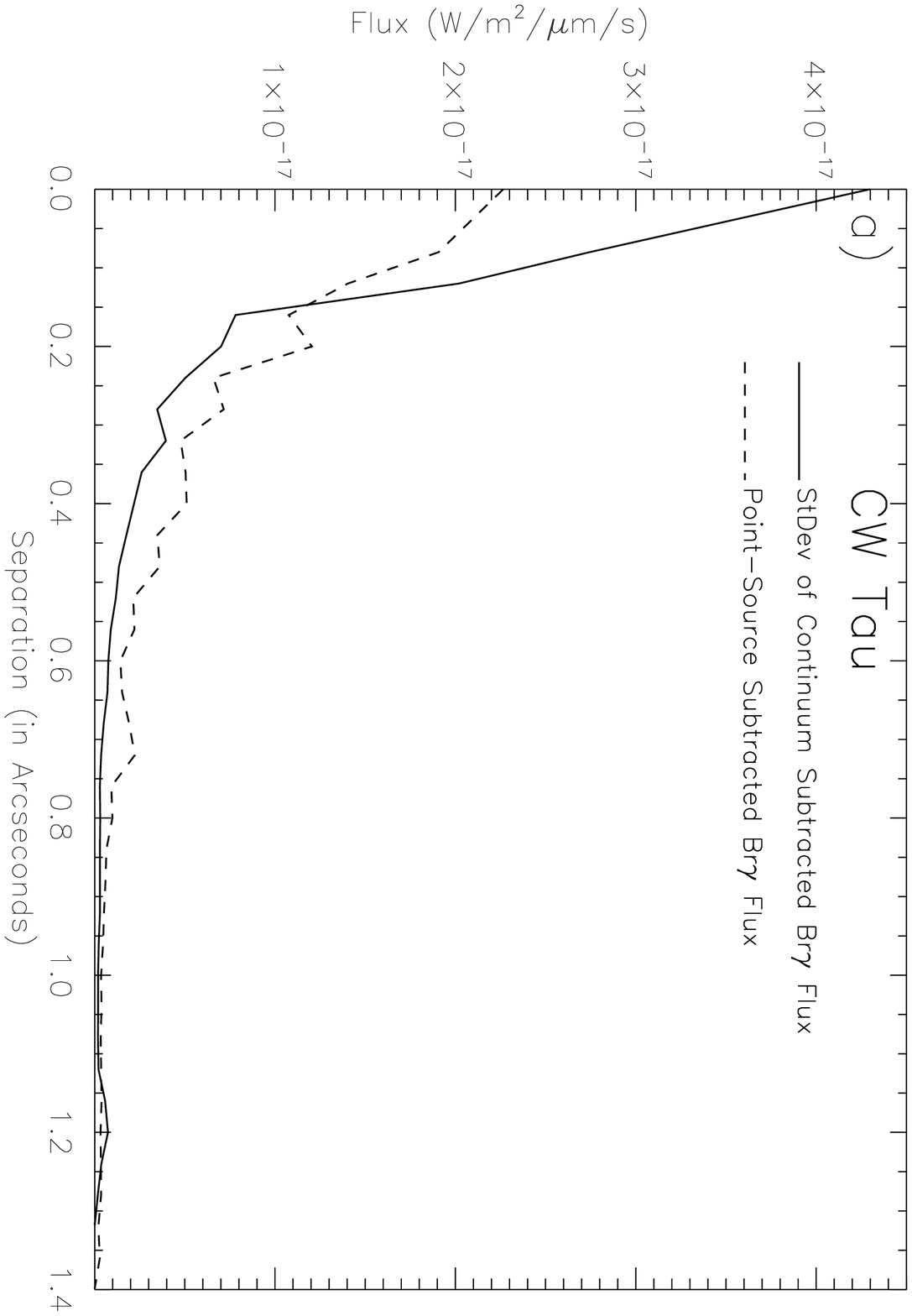}{350pt}{90pt}{40pt}{40pt}{20pt}{100pt}
\plotfiddle{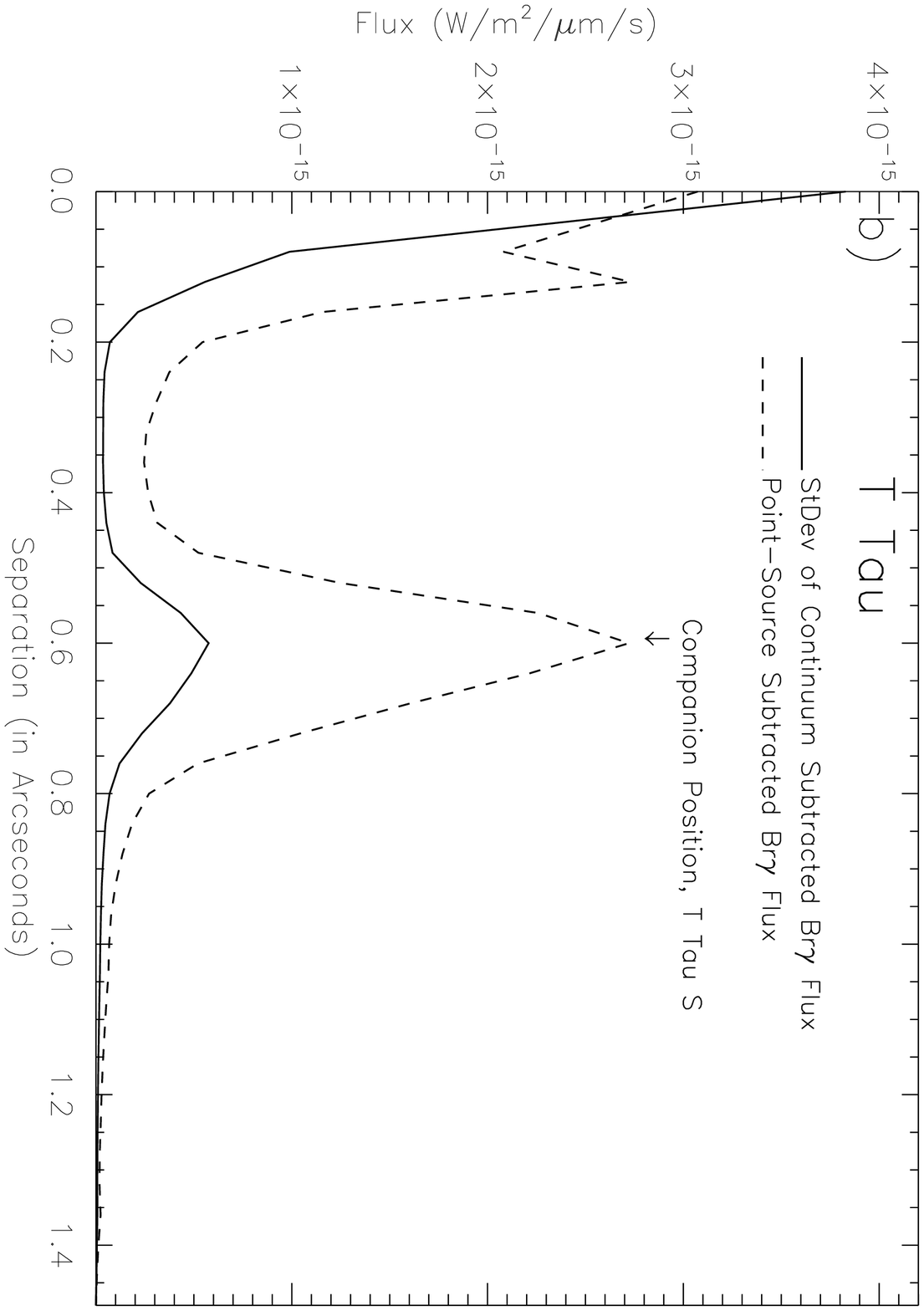}{350pt}{90pt}{40pt}{40pt}{300pt}{466pt}
\plotfiddle{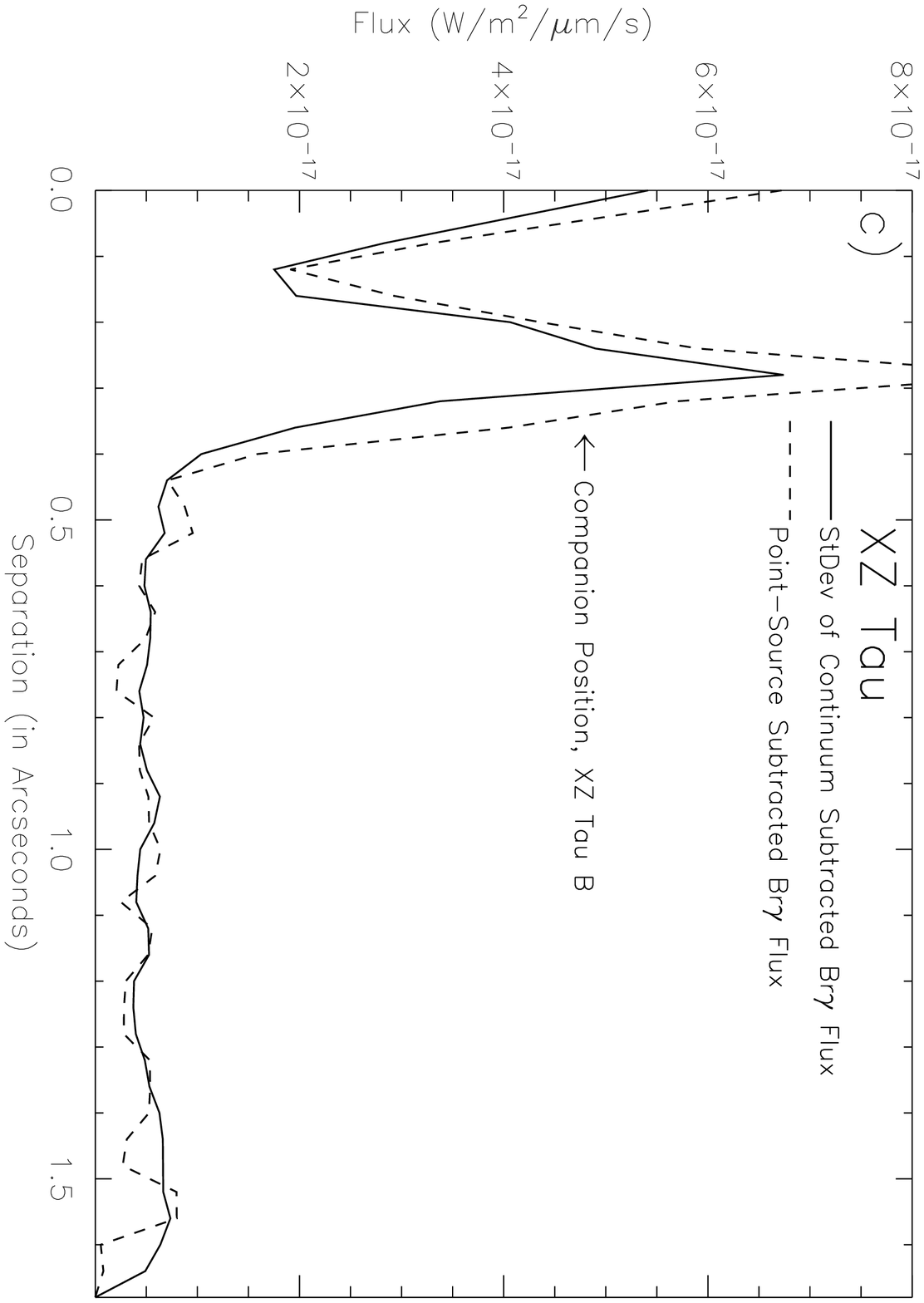}{350pt}{90pt}{40pt}{40pt}{20pt}{611pt}
\plotfiddle{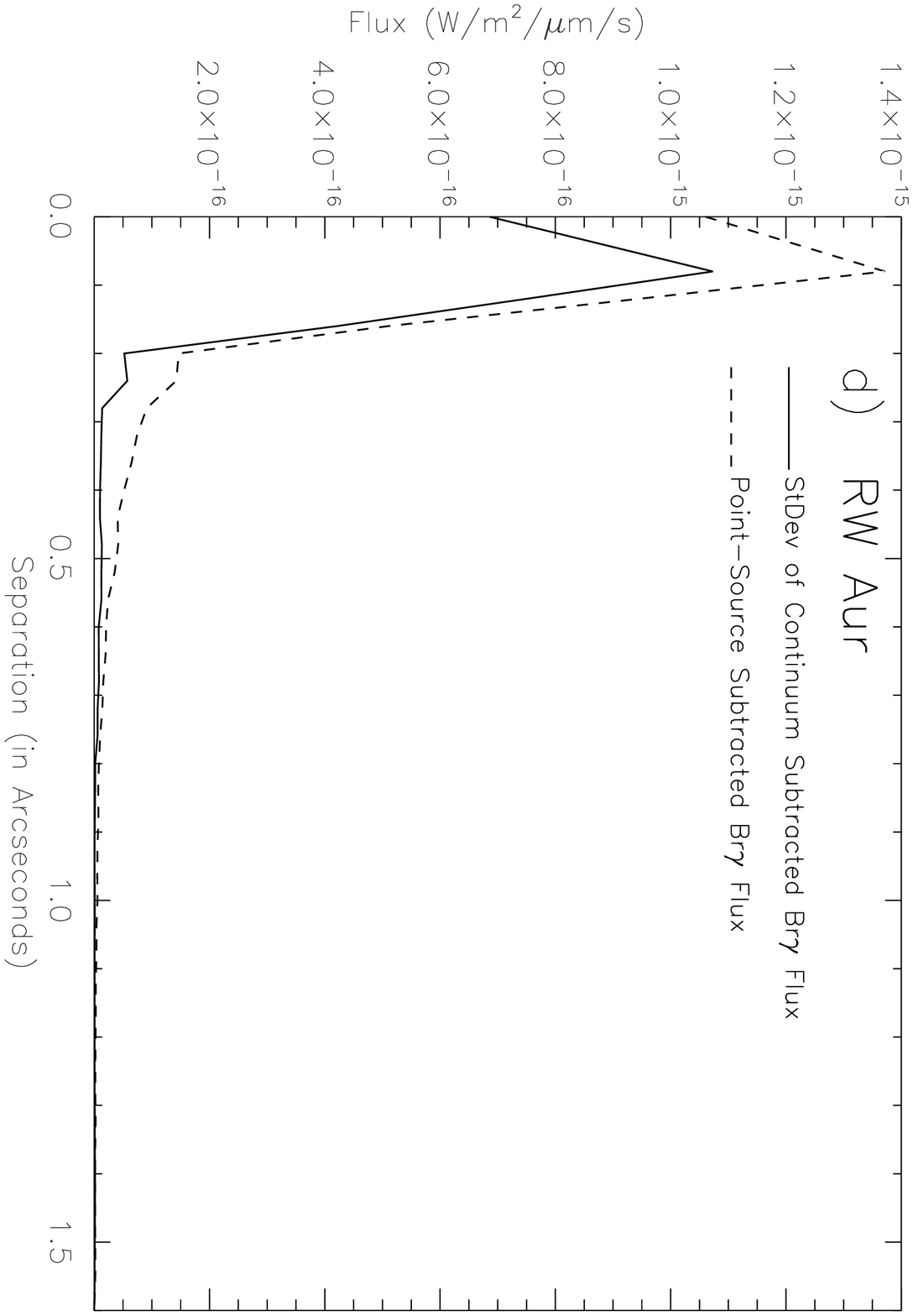}{350pt}{90pt}{40pt}{40pt}{300pt}{978pt}
\vskip -14.0in
\caption{The standard deviation of the continuum subtracted Br$\gamma$ flux plotted versus distance from the central continuum peak for CW Tau (a), T Tau (b), XZ Tau (c) and RW Aur (d).  The overplotted dashed line shows the signal from point-source subtracted Br$\gamma$ flux with increasing distance from the central star (e.g., XZ Tau A, RW Aur A, and T Tau N, in the case of the multiples).   Flux peaks at 0.$"$65 and 0.$"$25 seen in T Tau and XZ Tau represent the positions of the stellar companions in these systems.  The ratio of the dashed line to the solid line provides a measure of signal-to-noise in spatially extended Br$\gamma$ flux, though the companions complicate the detection.  It is clear from these plots that XZ Tau B and T Tau S have significant Br$\gamma$ emission above their continuum flux.  In these four cases, the S/N of spatially extended Br$\gamma$ is less than $\sim$3 and there is no convincing evidence for spatially extended emission in any of these TTSs.} 
\end{figure}

\end{document}